\documentclass[letterpaper,aps,twocolumn,nofootinbib,prl]{revtex4}
\usepackage[latin9]{inputenc}
\usepackage{color}
\usepackage{amsmath}
\usepackage{amssymb}
\usepackage{graphicx}
\usepackage{pifont}
\usepackage[english]{babel}
\usepackage{braket}
\usepackage{bm}
\usepackage[colorlinks,urlcolor=blue,citecolor=blue,linkcolor=blue]{hyperref}
\usepackage{soul}
\begin{document}

\title{Tuning anomalous Floquet topological bands with ultracold atoms}

\author{Jin-Yi Zhang$^{1,2,3}$}
\author{Chang-Rui Yi$^{1,2,3}$}
\author{Long Zhang$^{8,4,5}$}
\author{Rui-Heng Jiao$^{1,2,3}$}
\author{Kai-Ye Shi$^{6}$}
\author{Huan Yuan$^{1,2,3}$}
\author{Wei Zhang$^{6}$}
\author{Xiong-Jun Liu$^{4,5,7}$}
\author{Shuai Chen$^{1,2,3}$}
\author{Jian-Wei Pan$^{1,2,3}$}

\affiliation{$^{1}$Hefei National Research Center for Physical Sciences at the Microscale and School of Physical Sciences, University of Science and Technology of China, Hefei 230026, China\\
$^{2}$Shanghai Research Center for Quantum Science and CAS Center for Excellence in Quantum Information and Quantum Physics, University of Science and Technology of China, Shanghai 201315, China\\
$^{3}$Hefei National Laboratory, University of Science and Technology of China, Hefei 230088, China\\
$^{4}$International Center for Quantum Materials, School of Physics, Peking University, Beijing 100871, China\\
$^{5}$Collaborative Innovation Center of Quantum Matter, Beijing 100871, China\\
$^{6}$Department of Physics, Renmin University of China, Beijing 100872, China\\
$^{7}$Shenzhen Institute for Quantum Science and Engineering, Southern University of Science and Technology, Shenzhen 518055, China\\
$^{8}$School of Physics and Institute for Quantum Science and Engineering, Huazhong University of Science and Technology, Wuhan 430074, China
}

\date{\today}

\begin{abstract}
The Floquet engineering opens the way to create new topological states without counterparts in static systems.
Here, we report the experimental realization and characterization of new anomalous topological states with high-precision Floquet engineering for ultracold atoms trapped in a shaking optical Raman lattice.
The Floquet band topology is manipulated by tuning the driving-induced band crossings referred to as band inversion surfaces (BISs), whose configurations fully characterize the topology of the underlying states.
We uncover various exotic anomalous topological states by measuring the configurations of BISs which correspond to the bulk Floquet topology.
In particular, we identify an unprecedented anomalous Floquet valley-Hall state that possesses anomalous helical-like edge modes protected by valleys and a chiral state with high Chern number.
\end{abstract}

\maketitle
%%%%%%%%%%%%
Over one century ago, Gaston Floquet established a formalism to describe time-periodic systems~\cite{Floquet1883}, which, now called the Floquet theory~\cite{Shirley1965}, has advanced the very broad studies of non-equilibrium physics in quantum systems and played a vital role in the recent developments of simulating exotic quantum phases out of equilibrium~\cite{Eckardt2017, Oka2019, Rudner2020, Harper2020}. Periodical driving introduces an extra degree of freedom in time domain, which extends the characterization of equilibrium quantum phases to dynamically steady states. Such a generalization leads to new concepts which attract considerable attention in the recent years, such as time crystals~\cite{Sacha2017, Else2020} and Floquet topological phases~\cite{Lindner2011, Rechtsman2013, Jotzu2014, Flaschner2016, Rudner2013, Nathan2015, Roy2017, Liu2019, Wintersperger2020}.
The latter is a novel extension of the widely explored topological quantum phases in equilibrium systems~\cite{Hasan2010,Qi2011}, of which the comprehensive classification and characterization have been established~\cite{Kitaev2009,Chiu2016}.

Floquet engineering provides a versatile tool to manipulate topological phases beyond the realm of conventional solid-state counterparts~\cite{Rudner2013,Nathan2015,Roy2017,Yao2017}.
With coherent temporal control, the driven system not only can be tailored properly to realize the Hamiltonian of novel static systems such as the Haldane model~\cite{Jotzu2014}, but also may be exploited to create new states unreachable in static configurations.
For instance, an anomalous Floquet topological state has been realized experimentally in periodically-modulated hexagonal lattices \cite{Wintersperger2020}.
In general, a spatially $d$-dimensional ($d$D) Floquet phase may acquire nontrivial topology in the higher $(d+1)$-dimensional space-time domain. This renders a novel paradigm called anomalous Floquet topological phase, which hosts robust boundary modes that are unlinked to the bulk topology of $d$D Floquet bands, hence breaking down the conventional bulk-boundary correspondence. Conventionally such phases can be characterized with topology defined in the full $(d+1)$D space-time dimension~\cite{Rudner2013}, with broad classes of anomalous Floquet phases having been predicted~\cite{Nathan2015,Roy2017}, albeit many necessitates experimental discovery.
Recently, a theory~\cite{Zhang2020} shows that the anomalous Floquet topological phases can also be fully characterized by sub-dimensional topology defined on a momentum subspace called band inversion surfaces (BISs) of $d$D Floquet bands~\cite{Zhang2018}, without involving the complex evolution in time dimension. This characterization offers a powerful indicator to calibrate the engineering of novel Floquet topological phases.

Here, we experimentally realize, manipulate, and detect new anomalous Floquet topological states for ultracold atoms via a systematic BIS engineering.
By periodically driving a quantum anomalous Hall (QAH) model with 2D spin-orbit coupling, we control and observe the appearance of BISs as they successively appear in the Brillouin zone, with which various anomalous Floquet topological states corresponding to distinct BIS configurations are observed. Particularly, we realize a novel anomalous Floquet valley-Hall state, which features helical-like edge states at boundaries, and is in stark difference with the widely investigated valley-Hall effect in static solid materials~\cite{Gorbachev2014,Mak2014,Schaibley2016,Ren2016}, as well as unlike the anomalous Floquet state in periodically-modulated hexagonal lattices \cite{Wintersperger2020}.

\begin{figure}
  \centering
  \includegraphics[width=1\linewidth]{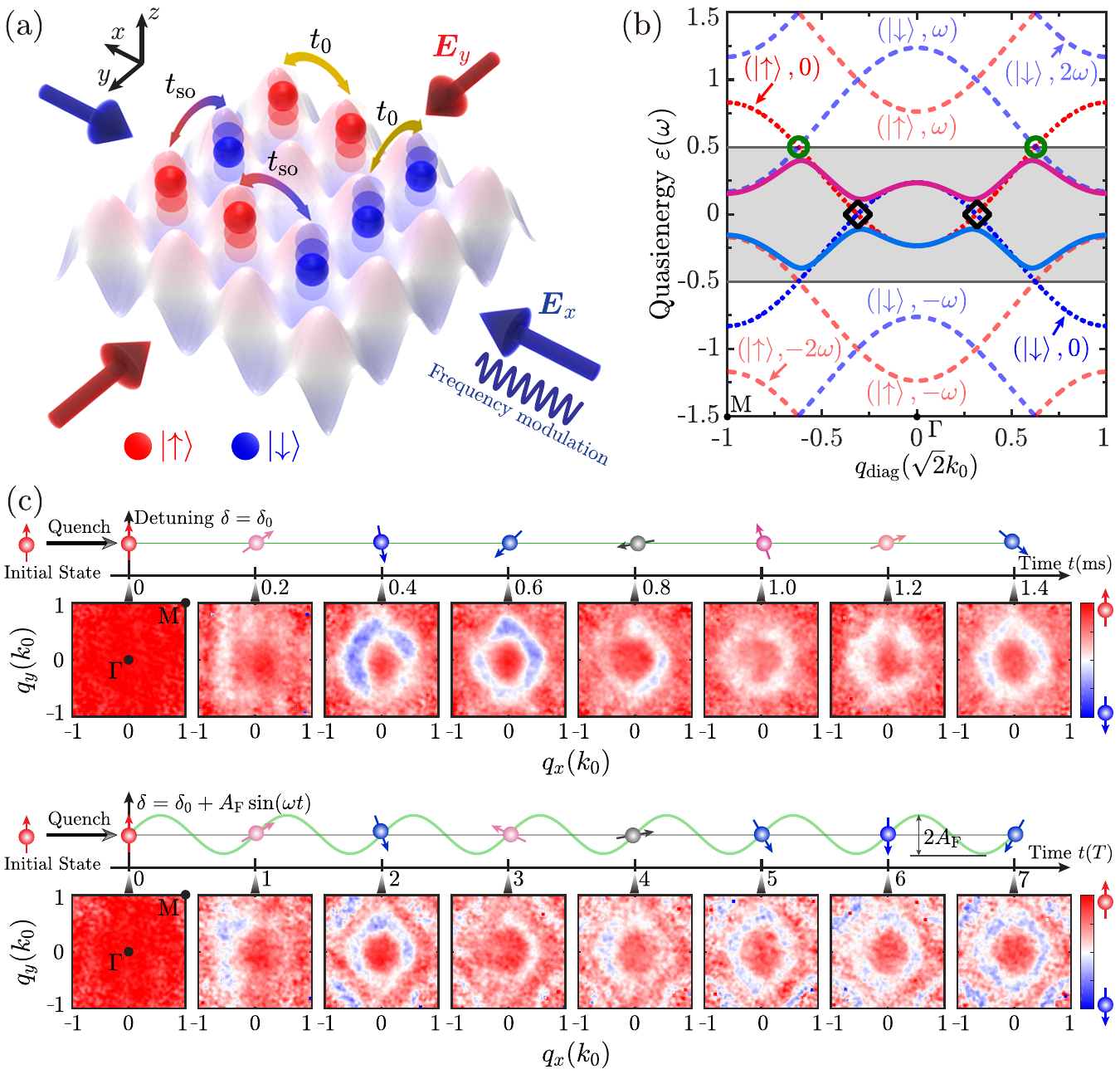}\\
  \caption{Scheme of shaking Raman lattices.
    (a) Cartoon of the optical Raman lattices with frequency modulation of a laser beam.
    (b) Sketch of the Floquet bands along the diagonal direction of the quasi-momentum $q_{\rm{diag}}$.
    $k_0$ is the wave number of the lattice beams, and $(\left| \uparrow \right\rangle,n\omega)$ ($(\left| \downarrow \right\rangle,n\omega)$) is the shifted copy of the static band $(\left| \uparrow \right\rangle,0)$ ($(\left| \downarrow \right\rangle,0)$).
    Solid curves are the Floquet bands which presenting 0-gap at the center of FBZ (black diamonds) and $\pi$-gap at the boundary of FBZ (green circles).
    (c) Experimental protocol and evolution of the spin textures $\left\langle \sigma_z(\bm{q},t) \right\rangle_z$.
    The upper (lower) row is spin textures without (with) periodic driving where the parameters $(V_0,~\Omega_0,~\delta_0)=(4.0,~1.0,~0.4)E_{\text{r}}$ ($(V_0,~\Omega_0,~\delta_0,A_{\text{F}})=(4.0,~1.0,~0.4,~0.8)E_{\text{r}}$, $T=400\mu \text{s}$).
    Here, $E_{\rm{r}}\approx h \times 3.7$kHz is the recoil energy.
    $\Gamma$ and M is high symmetric momenta.
    The red (blue) color in the spin textures denotes spin up (down).
  }\label{Fig1}
\end{figure}

Our experimental setup is a 2D periodically driven QAH model realized with ultracold $^{87}$Rb atoms trapped in shaking optical Raman lattices~\cite{SupMat} [see Fig.~\ref{Fig1}(a)].
The Raman lattices are conducted with magnetic sublevels $\left|F=1,m_F=-1\right\rangle$ (spin-up $\mid\uparrow\rangle$) and $\left|1,0\right\rangle$ (spin-down $\mid\downarrow\rangle$) in the $F=1$ manifold coupled by two Raman coupling processes~\cite{Wu2016,Sun2018a,Sun2018,Yi2019}.
Under the tight-binding approximation, the time-dependent Bloch Hamiltonian reads
\begin{align}~\label{eq1}
\begin{split}
\mathcal{H}(\bm{q},t)&=2t_{\text{so}}\sin{q_y}\sigma_x+2t_{\text{so}}\sin{q_x}\sigma_y\\
&+[\delta(t)/2-2t_0(\cos{q_x}+\cos{q_y})]\sigma_z,
\end{split}
\end{align}
where ${\bm q}=(q_x,q_y)$ and $\sigma_i$ are respectively the Bloch momentum and Pauli matrices $\bm{\sigma}$ acting on spin space, and the lattice constant is taken as unit of length.
The spin-conserved (spin-flip) hopping coefficient $t_0$ ($t_{\text{so}}$) is induced by the lattice (Raman) potential and set by the lattice depth $V_0$ (Raman strength $\Omega_0$) \cite{SupMat}.
The two-photon detuning is tuned in a form of a sinusoidal function $\delta(t)=\delta_0+A_{\text{F}}\sin\omega t$, which is realized by periodically modulating the laser frequency (Fig.~\ref{Fig1}(a)). Here, $\delta_0$ is a constant, $A_{\text{F}}$ denotes the modulation amplitude, and the driving frequency $\omega=2\pi/T$ with $T$ being the period.

Using the standard Floquet theory, this periodically driven QAH system can be delineated by an effective time-independent Hamiltonian $\mathcal{H}_{\text{F}}=\bm{h_\text{F}} \cdot \bm{\sigma}=h_{\text{F},x} \sigma_x+h_{\text{F},y}\sigma_y+h_{\text{F},z}\sigma_z$~\cite{Zhang2020}, whose eigen-energies are referred as quasienergies $\varepsilon=\pm |\bm{h}_{\text{F}}|$~(More details see \cite{SupMat}). Owing to the temporal periodicity, the quasi-energy band structure can be defined and tuned via the driving period $T=2\pi/\omega$.
Physically, the periodic modulation at frequency $\omega$ can transfer an energy in portions of $n\omega$ ($n$ is an integer) to the non-driven system (corresponding to $A_\text{F}=0$), such that the engineered Floquet band structure is formed by copying and shifting the static $(\mid\uparrow\rangle,0)$ and $(\mid\downarrow\rangle,0)$ bands within the Floquet Brillouin zone (FBZ) $\varepsilon \in [-\omega/2,~+\omega/2]$~\cite{Rudner2013,Zhang2020}. At the band crossing points, a finite $A_\text{F}$ and $\Omega_0$ will lift the degeneracy and open gaps, which can be further distinguished as $0$-gaps and $\pi$-gaps as they reside respectively at the center and the boundary of FBZ.
An example of Floquet band structure is shown in Fig.~\ref{Fig1}(b).
Remarkably, two types of BISs can be identified at the band crossing, on which $h_{\text{F},z}({\bf q})=0$~\cite{Zhang2020,SupMat}. While the conventional $0$-gap BIS (or $0$-BIS for short) is located within the 0-gap as in a static QAH model, an extra $\pi$-BIS is induced by driving within the $\pi$-gap as an exclusive feature of periodically driven systems.

\begin{figure}
  \centering
  \includegraphics[width=1\linewidth]{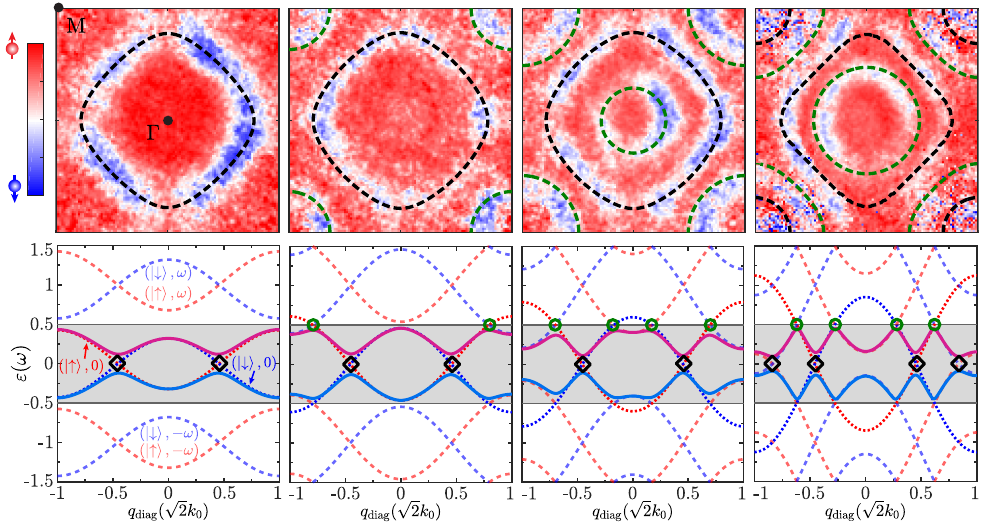}\\
  \caption{Tunable topological states.
    The emergence of ring patterns from one to four in the measured (first row) spin textures $\left\langle \sigma_z(\bm{q},t) \right\rangle$ and the Floquet bands along $q_{\rm{diag}}$ as $T$ varies with parameters $(V_0,~\Omega_0,~\delta_0)=(4.0,~1.0,~0.1)E_{\text{r}}$.
    And $(T,A_{\rm{F}},t)=(280\mu s,0.8E_{\rm{r}},2T)$, $(400\mu s,0.8E_{\rm{r}},3T)$, $(520\mu s,0.8E_{\rm{r}},2T)$, $(740\mu s,1.2E_{\rm{r}},2T)$ from left to right.
    The dashed black (green) curves mark the 0-BISs ($\pi$-BISs).
  }\label{Fig2}
\end{figure}

In this study, we synthesize unconventional Floquet topological bands by engineering the BIS configurations systematically and identify their topology by quench dynamics with the following protocol~\cite{Sun2018,Yi2019}. The $^{87}$Rb atoms are initially prepared in the $\mid\uparrow\rangle$ state at a temperature around 100nK and the initial detuning of $-200E_{\rm{r}}$. At the beginning of the experimental sequence, the detuning is quenched to the modulated value $\delta(t)$ within 200ns and $\delta_0 \in [-1,1]E_{\rm{r}}$ (Fig.~\ref{Fig1}(c)).
Thereupon, atoms are driven out of equilibrium and evolve under the post-quench time-dependent Hamiltonian Eq.~(\ref{eq1}).
After periodically modulating $\delta(t)$ for a certain time $t=nT$, the number of atoms of $\mid\uparrow\rangle$ ($N_{\uparrow}$) and $\mid\downarrow\rangle$ ($N_{\downarrow}$) at quasi-momentum $\bm{q}$ are measured by spin-resolved time-of-flight (TOF) imaging, and the dynamics of spin polarization $\left\langle\sigma_z(\bm{q},t)\right\rangle_{z}=[N_{\uparrow}(\bm{q},t)-N_{\downarrow}(\bm{q},t)]/[N_{\uparrow}(\bm{q},t)+N_{\downarrow}(\bm{q},t)]$ is extracted.
Here the subscript $z$ of $\langle\cdot\rangle_z$ denotes quenching $\delta$ at the beginning, corresponding to a process of quenching $h_{{\rm F},z}$ \cite{Zhang2018,Yi2019}. The BISs can be determined as the locations where full spin-flip takes place during such spin dynamics, and work as the key feature to characterize the topology~\cite{Sun2018,Zhang2018}.
As an example, we show the spin textures $\left\langle \sigma_z(\bm{q},t=n T) \right\rangle_z$ in the lower panel of Fig.~\ref{Fig1}(c), in comparison with the upper panel where no periodic driving is applied.
In the upper panel, there exists only one ring surrounding $\Gamma$ point of the Brillouin zone, corresponding to the BIS formed by the spin-orbit coupling of non-driven QAH model~\cite{Sun2018}. The periodic driving induces an additional ring around the $\text{M}$ point, indicating the emergence of a driving-induced BIS in the $\pi$-gap (Fig.~\ref{Fig1}(b)).

\begin{figure}
  \centering
  \includegraphics[width=1\linewidth]{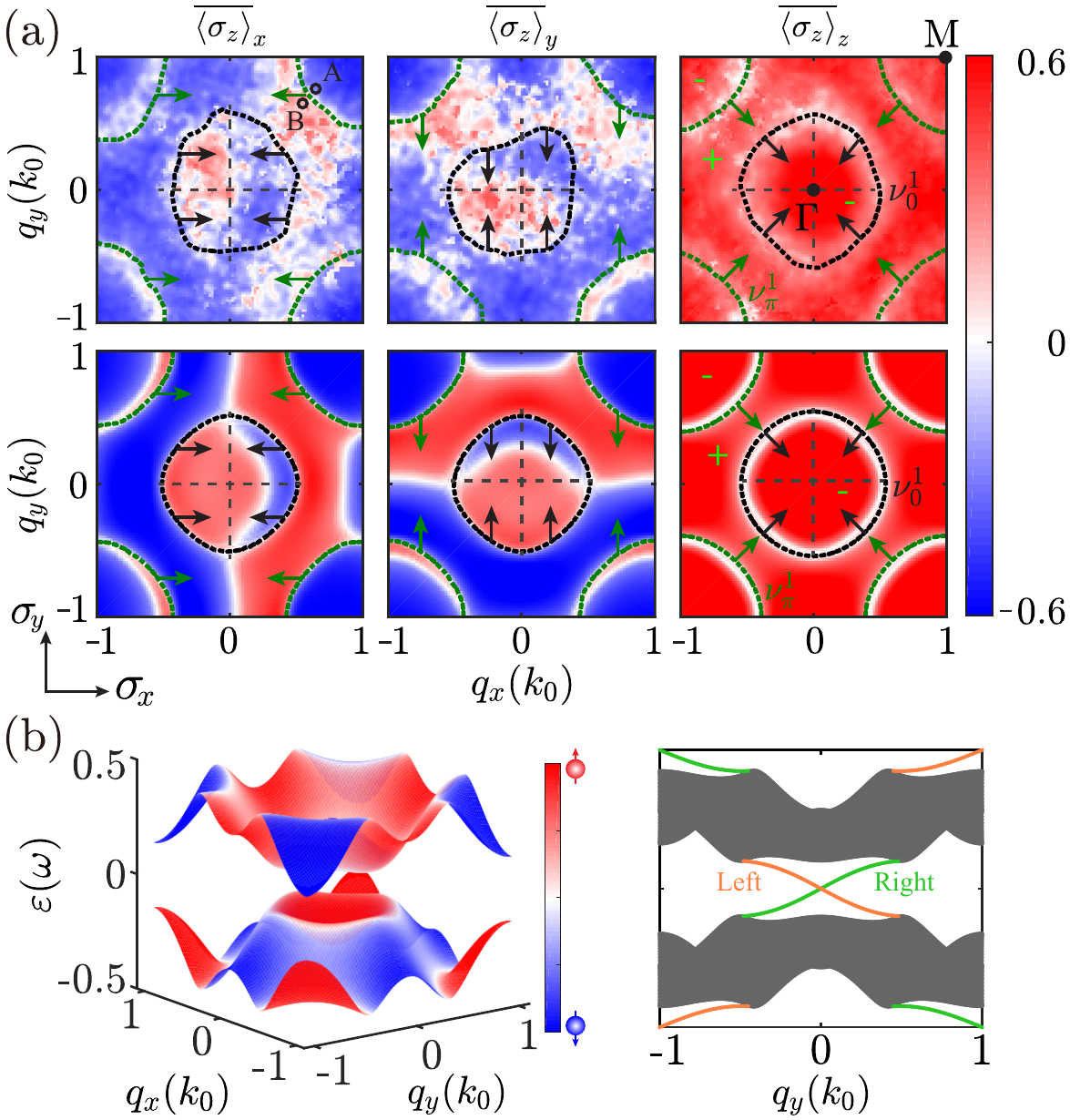}\\
  \caption{Characterization of topological states.
  (a) Time-averaged spin textures $\overline{\left\langle \sigma_z(\bm{q},t) \right\rangle}_{i}$ obtained in the experimental measurements (upper) and numerics with parameters in experiment (lower).
  The black and green curves represent the location of BISs.
  The dashed curves in the upper row are the guides to the eye.
  The arrows along the BISs display the dynamical field.
  The BISs divide the Brillouin Zone into three regions: two of them with $h_{\text{F},z}<0$ as labeled by ``$-$", and one of them with $h_{\text{F},z}>0$ as denoted by ``$+$".
  $\nu_0^j$ ($\nu_{\pi}^j$) is the topological number of the $j$-th BIS in 0-gap ($\pi$-gap).
  (b) The numerical band structures. The orange (green) curves denote edge states in the left (right) boundary. The parameters are ($V_0,\Omega_0,\delta_0,A_\text{F},T$)=($4.0E_\text{r},1.0E_\text{r},0.4E_\text{r},0.8E_\text{r},400\mu \text{s}$).
  }\label{Fig3}
\end{figure}

Now we demonstrate the scheme to realize novel Floquet topological states by engineering the BISs of Floquet bands, and identify the topology by measuring the BIS configurations.
The engineering and characterisation of a targeted topological state are implemented in two steps.
We first identify in which gap each BIS lives. This is achieved by monitoring the BISs as gradually increasing the period of the driving term applied to an initial determined static phase.
Then we measure by quench dynamics the topological invariant of each emerged BIS from the BIS circling configuration.

In Fig.~\ref{Fig2}, observed spin textures $\left\langle \sigma_z({\bm q},t)\right\rangle_z$ with different driving period are depicted in the first row.
For small $T$ (280$\mu$s), there exits only one BIS circling the $\Gamma$ point (with number of rings $\mathcal{N}_{\rm BIS}=1$), which is adopted from the static band structure and corresponds to the crossing of $(\mid\uparrow\rangle,0)$ and $(\mid\downarrow\rangle,0)$ bands in the FBZ at $\varepsilon=0$, corresponding to static 0-BIS.
The Floquet band structure is continuously connected to the static QAH bands, and the topology is not changed by the applied driving.
With increasing $T$, new BISs emerge in $0$-gap or $\pi$-gap due to the new band crossings between the
shifted copies of the static bands $(\mid\uparrow\rangle,n\omega)$ and $(\mid\downarrow\rangle,n\omega)$, and surround either the $\Gamma$ or M point.
There exists a BIS that maintains as $T$ increases, corresponding to the static 0-BIS.
Topological transitions thus occur with the open and close of gaps at the band crossing ring.
From the theoretical analysis, one concludes a property (P1): the two types of BISs always appear alternatively along the diagonal direction of quasi-momentum $q_{\rm diag}$ for the present system (More details see~\cite{SupMat}).
With property (P1) and the static 0-BIS, we precisely identify whether the BIS is a 0-BIS or a $\pi$-BIS.
In experiment, we tune $T$ up to $740 \mu \text{s}$, and encounter four different regions with $\mathcal{N}_{\rm{BIS}}$ up to 4.

To fully characterize the underlying topology by the configurations of BISs, $\mathcal{N}_{\rm{BIS}}$ alone is not enough, and an extra set of topological invariants $\nu_{0,\pi}^{j}$ associated with the $j$-th BIS have to be employed to determine the winding number of the corresponding gap as $\mathcal{W}_{0,\pi} = \sum_{j} \nu_{0,\pi}^{j}$~\cite{Zhang2020}.
The value of $\nu_{0,\pi}^{j}$ can be extracted by the dynamical field, defined as the variation of time-averaged spin textures $\overline{\left\langle \sigma_z(\bm{q}) \right\rangle}_{i=x,y,z}$ across the BIS. Such $\overline{\left\langle \sigma_z(\bm{q}) \right\rangle}_{i}$ is obtained by quenching $h_{\text{F},i}$~\cite{Zhang2020,Yi2019,SupMat}.
Fig.~\ref{Fig3}(a) shows $\overline{\left\langle \sigma_z(\bm{q}) \right\rangle}_{x,y,z}$ and the corresponding numerical calculation for ${\cal N}_{\rm BIS}=2$.
Two BISs are identified as two rings circling the $\Gamma$ and $\text{M}$ points, respectively.
Based on the sequence of their emergence, we determine that the inner BIS surrounding the $\Gamma$ point corresponds to the static $0$-BIS, while the outer one is a driving-induced $\pi$-BIS.
These two BISs divide the Brillouin zone into three regions with $h_{\text{F},z}>0$ or $h_{\text{F},z}<0$, which can be judged by the properties of band inversions~\cite{SupMat}.
The two-component dynamical field is obtained by calculating the gradient from the region ``$-$" to ``+" (More details see~\cite{SupMat}).
From the top right corner of the BIS surrounding M point in spin texture $\overline{\left\langle \sigma_z \right\rangle}_x$, the difference of the spin polarisation between region A and region B is negative, hence the direction of the $x$-component dynamical field is along the opposite direction of $\sigma_x$.
The $y$-component dynamical field is obtained similarly from $\overline{\left\langle \sigma_z \right\rangle}_y$.
The combined dynamical field on BISs is then determined by superimposing the two components and depicted as arrows in
$\overline{\left\langle \sigma_z \right\rangle}_{z}$, which gives a nonzero $\nu_{0,\pi}^{j}$ for each BIS in the $0$ and $\pi$ gaps.
In this case, we get $\nu_0^{1}=+1$ for the 0-BIS surrounding the $\Gamma$ point and $\nu_{\pi}^{1}=-1$ for the $\pi$-BIS circling the $\text{M}$ point (More details see~\cite{SupMat}).
The winding numbers are then $\mathcal{W}_0=\nu_0^{1}=+1$ and $\mathcal{W}_{\pi}=\nu_{\pi}^{1}=-1$, respectively, showing a high Chern number of the anomalous Floquet band as $\mathcal{C}=\mathcal{W}_0-\mathcal{W}_{\pi}=+2$.
As expected, the Chern number of the present anomalous Floquet topological band does not directly correspond to the edge states of each gap under open boundary condition (Fig.~\ref{Fig3}).
Instead, the chiral edge states are directly determined by $\nu_{0,\pi}^{j}$
(see the right panel of Fig.~\ref{Fig3}(b) and ~\cite{SupMat}).

\begin{figure}
  \centering
  \includegraphics[width=1\linewidth]{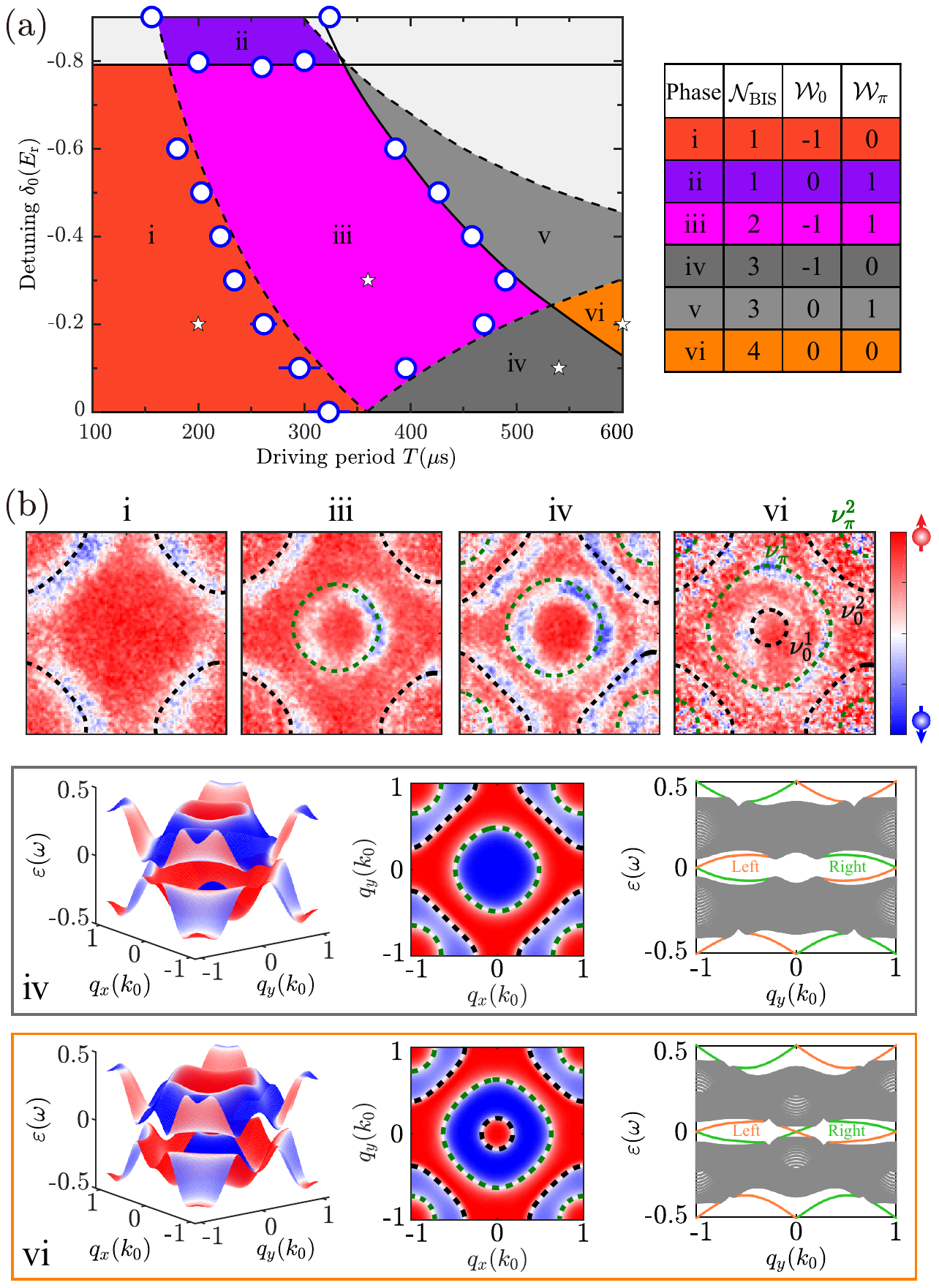}\\
  \caption{ Topological phase diagram.
    (a) Topological phase diagram with $(V_0, \Omega_0)=(4.0, 1.0)E_\text{r}$ in the $T$-$\delta_0$ plane.
    The blue circles with error bars are experimental measurements.
    The solid and dashed curves are calculated phase boundaries induced by 0-gap and $\pi$-gap, respectively.
    Roman numerals {\romannumeral1} to {\romannumeral6} label 6 topological phases within the range of experimental parameters, whose topological invariants are listed in the table.
    (b) Spin textures $\left\langle \sigma_z(\bm{q},t) \right\rangle_z$ marked by {\text{\ding{73}}} with typical phases in (a).
    To clearly observe {\romannumeral6}, the raw data are shown in Fig.~S8 of ~\cite{SupMat}.
    The BIS configurations labeled by dashed black and green curves fully distinguish the topology of all phases.
    The following gray (orange) frame contains the bulk band structures, the aerial view of the lowest band and the band structures with open boundaries corresponding to the spin texture {\romannumeral4} ({\romannumeral6}).
    The orange and green curves denote ``helical'' edge modes.
  }\label{Fig4}
\end{figure}

The above observation shows the three basic features which enable us to realize, manipulate and detect the anomalous Floquet topological states through a systematic engineering of the BIS configurations~\cite{SupMat,zhang2021}.
First, the precise location in the $0$-gap or $\pi$-gap of the alternatively emerging BISs is determined by gradually increasing the driving period $T$ and the property (P1).
Second, $\nu_{0,\pi}^{j}$ is uniquely determined by the BIS circling configuration.
Namely, $\nu_{0,\pi}^{j}$ is $+1$ ($-1$) when the BIS surrounds the $\Gamma$ ($\text{M}$) point (P2).
Third, each BIS with $\nu_{0,\pi}^{j}\ne 0$ corresponds to a chiral edge modes if the open boundary is considered (P3).
With these features we map out the rich phase diagram of the present system in Fig.~\ref{Fig4}(a) with $\delta_0<0$ and the complete phase diagram is shown in the supplementary materials~\cite{SupMat}.
The phase boundaries are determined by fitting the size of the rings in the spin textures as a function of $T$ or $\delta_0$~\cite{SupMat}.

The most striking topological state uncovered in the current experiment is the one labeled by ``vi'', in which four BISs are observed in the spin textures depicted in Fig.~\ref{Fig4}(b).
The inner two BISs surround the $\Gamma$ point and the outer two wind around the $\text{M}$ point.
Through the regulation of driving period (Fig.~\ref{Fig2}), we determine that the outer black ring is the static $0$-BIS, while the other three BISs alternatively lives in the 0-gap and $\pi$-gap according to the property (P1).
The circling configuration further determines
$\nu_0^1 = +1$, $\nu_{\pi}^1 = +1$, $\nu_0^2 = -1$ and $\nu_{\pi}^2 = -1$ from center to lattice Brillouin zone edge, leading to the vanishing winding numbers in each quasienergy gap $\mathcal{W}_0=\nu_0^{1}+\nu_0^{2}=0$ and $\mathcal{W}_{\pi}=\nu_{\pi}^{1} + \nu_{\pi}^2=0$, and Chern number $\mathcal{C} = 0$. However, from the BIS-boundary correspondence (P3) the four BISs correspond to two pairs of edge modes with opposite chirality residing in the $0$ and $\pi$ gaps, respectively, despite the trivial global topological invariants, consistent with the numerical results under the same parameter conditions in experiment (see the lowest panel of Fig.~\ref{Fig4}(b)).

The counter-propagating edge modes within each gap are protected by local topological structures (BISs) in two different Brillouin zone regions, i.e. the valleys, and are thus robust against long-range disorders which do not induce inter-valley scatterings~\cite{Ren2016,SupMat}, rendering an exotic {\it anomalous Floquet valley-Hall state} with multiple ``helical'' edge modes.
Such modes, similar to ``spin-helical'' edge modes in the quantum spin Hall effect, have a valley degree of freedom now coupled to the direction of motion (see Fig.~\ref{Fig4}(b)).
This topological state is beyond the classification and characterization of anomalous topological phases using global topological invariants of Floquet bands and winding number of each gap~\cite{Lababidi2014,Umer2020}, but is fully captured by the BIS--boundary correspondence.
A similar analysis on the phase labeled by ``iv'' with ${\cal N}_{\rm BIS}=3$ gives $\mathcal{W}_{0}=-1$, $\mathcal{W}_{\pi}=0$ and $\mathcal{C}=-1$, which protects counter-propagating edge modes inside the $\pi$-gap (the middle panel of Fig.~\ref{Fig4}(b)).

In conclusion, we experimentally implemented a systematic and powerful technique of spatiotemporal control to realize, engineer and detect novel anomalous Floquet topological states.
From quench measurement, we observe that the topology of an anomalous Floquet state in the space-time $(d+1)$-dimension uniquely corresponds to the nontrivial configuration of $(d-1)$-dimensional BISs.
This enables a highly feasible experimental approach to realize and engineer new Floquet topological states, including a novel anomalous valley-Hall state, which is a new paradigm achieved here and beyond the existing anomalous Floquet phases~\cite{Rudner2013,Kitagawa2012,Hu2015,Maczewsky2017,Mukherjee2017,Wintersperger2020}.
The technique can be directly applied to fermionic systems and extended to high dimensions~\cite{zhang2021}, and shall open up a broad avenue of space-time manipulation towards exploring new fundamental topological states of matter.

\begin{acknowledgments}
We acknowledge insightful discussions with Zong-Yao Wang, Xiang-Can Cheng, Wei Zheng and Hui Zhai. This work was supported by the Innovation Program for Quantum Science and Technology (Grant No. 2021ZD0302000), the National Key Research and Development Program of China (Grant No. 2021YFA1400900 and 2018YFA0306501), the National Natural Science Foundation of China (Grant No. 12025406, 11825401, 12074428 and 12104445), Anhui Initiative in Quantum Information Technologies (Grant No. AHY120000), Shanghai Municipal Science and Technology Major Project (Grant No. 2019SHZDZX01), and the Strategic Priority Research Program of Chinese Academy of Science (Grant No. XDB28000000). J.-Y.Z. acknowledges support from the startup grant of University of Science and Technology of China (Grant No. KY2340000152) and the Sponsored by Shanghai Pujiang Program (Grant No. 21PJ1413600). L. Z. acknowledges support from the startup grant of Huazhong University of Science and Technology (Grant No. 3004012191). W.Z. acknowledges support from the Beijing Natural Science Foundation (Grant No. Z180013). X.-J.L. acknowledges support from the Open Project of Shenzhen Institute of Quantum Science and Engineering (Grant No. SIQSE202003).
\end{acknowledgments}

%\bibliography{SOCFloquet}
%

\newpage
\onecolumngrid
\renewcommand\thefigure{S\arabic{figure}}
\setcounter{figure}{0}
\renewcommand\theequation{S\arabic{equation}}
\setcounter{equation}{0}
\makeatletter
\newcommand{\rmnum}[1]{\romannumeral #1}
\newcommand{\Rmnum}[1]{\expandafter\@slowromancap\romannumeral #1@}
\makeatother

\newpage

{
\center \bf \large
Supplemental Material for: \\
Tuning anomalous Floquet topological bands with ultracold atoms\vspace*{0.1cm}\\
\vspace*{0.0cm}
}

\vspace{4ex}

\maketitle

\subsection{
 2D optical Raman lattices for ultracold Bose gases}

\begin{figure}
  \centering
  \includegraphics[width=0.5\linewidth]{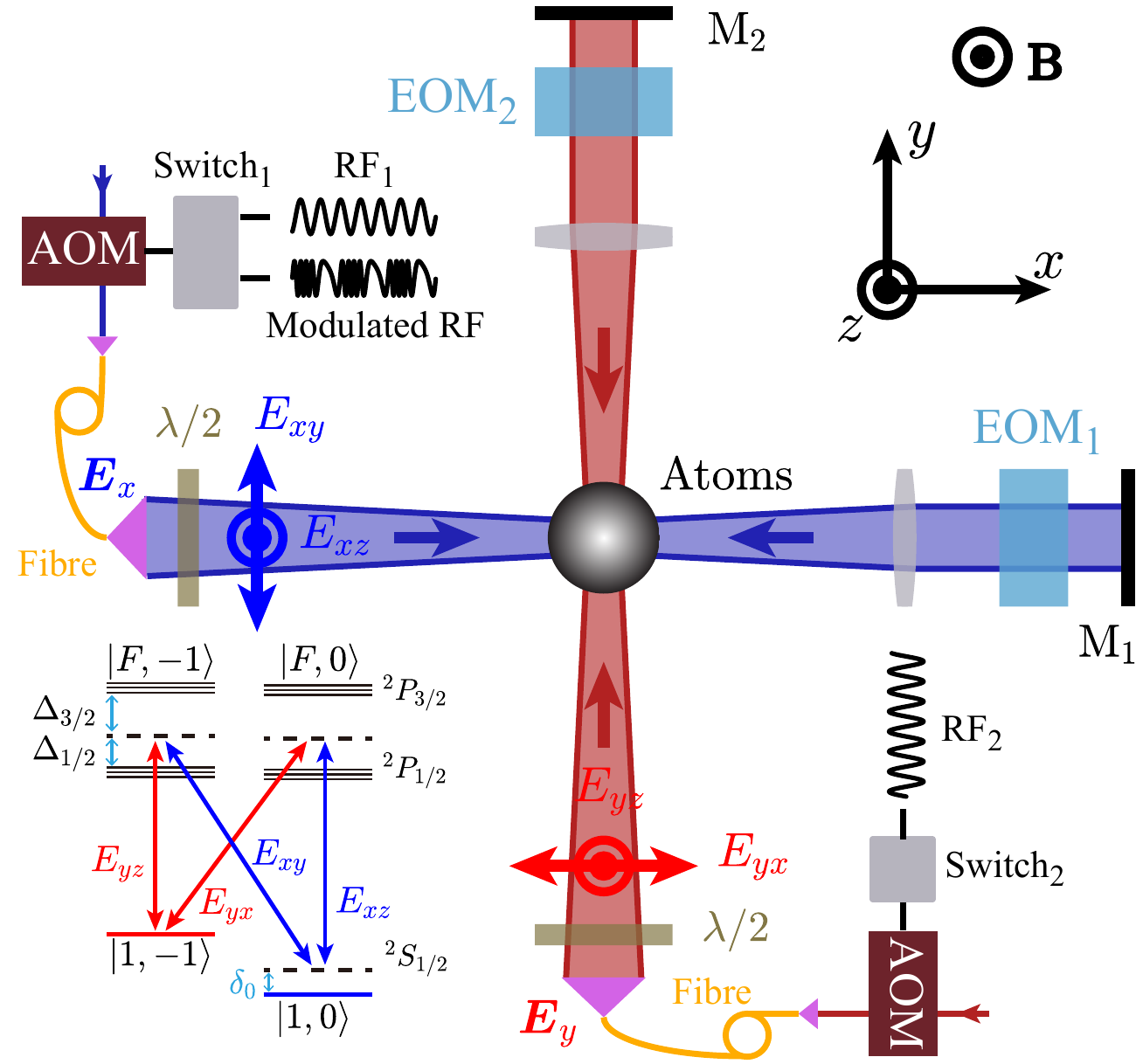}
  \caption{Experimental setup.
   $\bm{E}_x$ ($\bm{E}_y$) represents the laser beam, possessing two orthogonally polarised components $\bm{E}_{xy}$ $(\bm{E}_{yx})$ and $\bm{E}_{xz}$ $(\bm{E}_{yz})$. $\lambda/2$ represents the half-wave plate; $\text{M}_1$ ($\text{M}_2$) denotes the mirror; $\text{EOM}_{1,2}$ are the electro-optic modulators; AOM is the acousto-optic modulator; $\text{RF}_{1,2}$ are the ratio-frequency signals; $\text{Switch}_{1,2}$ are RF switches. Bias magnetic field $\textbf{B}$ is along $z$ direction. Inset: level structures and Raman couplings.
  }\label{FigS1}
\end{figure}

The experimental setup is demonstrated in Fig.~\ref{FigS1}.
Ultracold atomic cloud of $^{87}$Rb is prepared in a dipole optical trap.
Laser beam $\bm{E}_x$ ($\bm{E}_y$) with wavelength $\lambda=787$nm along $\hat{x}$ ($\hat{y}$) direction is incident on the clouds, then it is reflected back to atomic clouds by the mirror $\text{M}_1$ ($\text{M}_2$), passing through a $\lambda/2$ wave plate and electro-optic modulator (EOM).
The $\lambda/2$ wave plate splits the beam $\bm{E}_{x}$ ($\bm{E}_y$) into two orthogonally polarised components $\bm{E}_{xy}$ $(\bm{E}_{yx})$ and $\bm{E}_{xz}$ $(\bm{E}_{yz})$, and the $\text{EOM}_1$ ($\text{EOM}_2$) induces the phase shift $\varphi_1$ $(\varphi_2)$ between the two components $\bm{E}_{xy}$ and $\bm{E}_{xz}$ $(\bm{E}_{yx}$ and $\bm{E}_{yz})$.
Moreover, two magnetic sublevels $\mid\uparrow\rangle\equiv\left|F=1,m_F=-1\right\rangle$ and $\mid\downarrow\rangle\equiv\left|1,0\right\rangle$ are subjected to a bias magnetic field $\textbf B$ along $\hat{z}$ direction, generating a Zeeman splitting of $10.2\text{MHz}$.
Thus, the squared optical lattice potentials $V_{\text{latt}}(x,y)$ and the Raman lattices $\Omega_{x,y}(x,y)$ can be tuned by $\varphi_{1,2}$~\cite{Yi2019}, i.e.,
\begin{equation}
\begin{aligned}
V_{\text{latt}}(x,y)&=V_{0}^{\prime}[|\bm{E}_{xz}|^2\cos^2(k_0x-\varphi_1)+|\bm{E}_{xy}|^2\cos^2(k_0x)+|\bm{E}_{yz}|^2\cos^2(k_0y-\varphi_2)+|\bm{E}_{yx}|^2\cos^2(k_0y)],\\ \Omega_{x}(x,y)&=\Omega_{0x}\cos (k_0x-\varphi_1)\cos (k_0y)\exp(-i\varphi_1), \Omega_{y}(x,y)=\Omega_{0y}\cos (k_0x)\cos (k_0y-\varphi_2)\exp(i\varphi_2).
\end{aligned}
\end{equation}
Here, $k_0=2\pi/\lambda$, $V_0^{\prime}$ is determined by the level structure of ${}^{87}$Rb atoms and the Raman coupling strength $\Omega_{0x,0y}\propto |\bm{E}_{yx,xy}||\bm{E}_{xz,yz}|$.
In the experiment, we set $|\bm{E}_{xy}|=|\bm{E}_{yx}|$ and $|\bm{E}_{xz}|=|\bm{E}_{yz}|$, then $\Omega_{0x}=\Omega_{0y}=\Omega_0$.
And the total Hamiltonian reads
\begin{equation}
H=\begin{pmatrix}
 \frac{\bm{p}^2}{2m}+V_{\text{latt}}(x,y)+\frac{\delta}{2}& \Omega_x-i\Omega_y\\
  \Omega_x+i\Omega_y&\frac{\bm{p}^2}{2m}+V_{\text{latt}}(x,y)-\frac{\delta}{2}
\end{pmatrix},
\end{equation}
where $\delta$ is the two-photon detuning.
With the tight-binding approximation, the above Hamiltonian generally can be written as
\begin{equation}\label{HamGeneral}
 {\cal H}_{0}(\bm q)=\bm{h}(\bm{q})\cdot\bm{\sigma}=(m_x+2t^x_{\rm so}\sin q_y)\sigma_x+(m_y+2t^y_{\rm so}\sin q_x)\sigma_y+[\delta/2-2(t^x_0\cos q_x+t_0^y\cos q_y)]\sigma_z.
\end{equation}
Here, the Zeeman constants $m_{x,y}$ and the spin-conserved (flipped) hopping coefficients $t_0^{x,y}$ ($t_{\rm so}^{x,y}$) are depends on $\varphi_{1,2}$.
The Pauli matrices $\sigma_{x,y,z}$ are defined as
\begin{equation*}
\sigma_x=\begin{pmatrix}
  0&1 \\
  1&0
\end{pmatrix},
\sigma_y=\begin{pmatrix}
  0&-i \\
  i&0
\end{pmatrix},
\sigma_z=\begin{pmatrix}
  1&0 \\
  0&-1
\end{pmatrix}.
\end{equation*}

When $\varphi_1=\varphi_2=\pi/2$, the squared optical lattice potentials $V_{\text{latt}}(x,y)$ and the Raman lattices $\Omega_{x,y}(x,y)$ can be written as
\begin{equation}
\begin{aligned}
\label{eqS1}
V_{\text{latt}}(x,y)&=V_{0x}\cos^2k_0x+V_{0y}\cos^2k_0y,\\
\Omega_{x}(x,y)&=\Omega_{0x}\sin k_0x\cos k_0y,\\
\Omega_{y}(x,y)&=\Omega_{0y}\cos k_0x\sin k_0y.
\end{aligned}
\end{equation}
where the lattice depth $V_{0x,0y}\propto\lvert \bm{E}_{xy,yx} \rvert^2-\lvert \bm{E}_{xz,yz} \rvert^2$.
Then, $V_{0x}=V_{0y}=V_0$ due to $|\bm{E}_{xy}|=|\bm{E}_{yx}|$ and $|\bm{E}_{xz}|=|\bm{E}_{yz}|$, and the Hamiltonian of the Raman lattices reads
\begin{equation}
\begin{aligned}
\label{eqS2}
H=\frac{\bm{p}^2}{2m}+V_{\text{latt}}(x,y)+\Omega_x\sigma_x+\Omega_y\sigma_y+\frac{\delta}{2}\sigma_z.
\end{aligned}
\end{equation}
With the tight-binding approximation, the Hamiltonian is given in the Eq.(1) of main text, i.e.,
\begin{equation}\label{HamTBM}
\mathcal{H}_0(\bm{q})=2t_{\text{so}}\sin{q_y}\sigma_x+2t_{\text{so}}\sin{q_x}\sigma_y+[\delta/2-2t_0(\cos{q_x}+\cos{q_y})]\sigma_z.
\end{equation}
Here, $m_x=m_y=0$, the spin-conserved hopping coefficients $t_{0}^x=t_{0}^y=t_{0}$ and spin-flipped hopping coefficients $t_{\rm{so}}^x=t_{\rm{so}}^y=t_{\rm{so}}$ are, respectively, %spin-conserved and
\begin{align}
t_0&=-\int d{\bm r}\phi_{s}(x,y)\left[\frac{{\bm k}^2}{2m}+V_{\rm latt}({\bm r})\right]\phi_{s}(x-1,y),\nonumber\\
t_{\rm so}&=\Omega_{0}\int d{\bm r}\phi_{s}(x,y)\cos(k_0y)\sin(k_0x)\phi_{s}(x-1,y),
\end{align}
where $\phi_{s}(x,y)$ denotes the Wannier function of the lowest band.
For the typical parameters $V_0=4E_{\rm{r}}$ and $\Omega_0=1E_{\rm{r}}$ in the Raman lattices, $t_0\approx0.09$ and $t_{\rm{so}}\approx 0.05$.

When ($\varphi_1,\varphi_2$)=($0,\pi/2$) or ($0,\pi/2$), the Bloch Hamiltonian in the tight-binding limit reads~\cite{Yi2019}
\begin{equation}
 {\cal H}_0=[\delta/2-2(t^x_0\cos q_x+t_0^y\cos q_y)]\sigma_z+2t^x_{\rm so}\sin q_y\sigma_x+m_y\sigma_y,
\end{equation}
or
\begin{equation}
 {\cal H}_0=[\delta/2-2(t^x_0\cos q_x+t_0^y\cos q_y)]\sigma_z+2t^y_{\rm so}\sin q_x\sigma_y+m_x\sigma_x,
\end{equation}
where the Zeeman constant
\begin{equation}
m_y=\Omega_{0x}\int d{\bm r}\phi_{s}({\bm r})\cos(k_0x)\cos(k_0y)\phi_{s}({\bm r})\gg t_{\rm so}^x, t_0^{x,y}
\end{equation}
or
\begin{equation}
m_x=\Omega_{0y}\int d{\bm r}\phi_{s}({\bm r})\cos(k_0x)\cos(k_0y)\phi_{s}({\bm r})\gg t_{\rm so}^y, t_0^{x,y}.
\end{equation}

The shaking Raman lattices are realized by modulating the two-photon detuning $\delta$.
To this end, the frequency of the laser beam $\bm{E}_x$ is modulated by a ratio-frequency (RF) signal which drives an acousto-optic modulators (AOM) (Fig.~\ref{FigS1}).
Meanwhile, the frequency of the laser beam $\bm{E}_y$ is fixed.
Thus, the two-photon detuning $\delta=\delta_0+A_\text{F}\sin{\omega t}$.

\subsection{Quench protocol for three quantized axes}
To better understand quench protocol, we shall first present the basic idea of quench dynamics that was described in the Ref.~\cite{Sun2018,Yi2019}.
Our system is prepared initially in a topological trivial state via controlling the detuning or the phase of the beams.
Thereupon, the system is quenched to the topological regime and the initial state evolves under the post-quench Hamiltonian.
Effectively, the quench projects the initial state to the eigenstates of the post-quench Hamiltonian, which are the superposition states of $\mid\uparrow\rangle$ and $\mid\downarrow\rangle$.
Therefore, under the post-quench Hamiltonian the state at each ${\bf q}$ oscillates between $\mid\uparrow\rangle$ and $\mid\downarrow\rangle$.
The oscillation frequency is governed by the energy difference of eigen-bands.
The quench dynamics is similar to the Rabi oscillation of a two-level system in quantum optics.

In experiments, we quench three quantized axes of the effective Hamiltonian $h_{{\rm F},i}$ ($i=x,y,z$) to explore and characterise the novel topological phases \cite{Zhang2018,Zhang2020,Sun2018,Yi2019}.
Quenching $h_{\text{F},z}$ is realised by fast switching the two-photon detuning $\delta$,
while quenching $h_{{\rm F},x}(h_{{\rm F},y})$ is performed by suddenly tuning the phase of Raman couplings $\varphi_2(\varphi_1)$.
For all the quenches, we divide them into three steps: the preparation of the initial states, quenching to the objective Hamiltonian and the detection of the states. Here, the objective Hamiltonian Eq.~\ref{HamTBM} possesses the parameters $V_0=4.0E_{\rm{r}}, \Omega_0=1.0E_{\rm{r}}$.
Quench protocol is performed as following:

(1) \emph{The preparation of the initial states.} The ${}^{87}$Rb atoms are prepared slightly above the critical temperature of Bose-Einstein condensation.
Subsequently, they are adiabatically loaded into the Raman lattices in 100ms with temperature of 100nK.
Regarding quenching $h_{{\rm F},z}$, we set the initial two-photon detuning $\delta=-200E_{\rm{r}}$ by tuning the frequency of $\text{RF}_1$ and the phase $(\varphi_1, \varphi_2)=(\pi/2, \pi/2)$.
Then, $h_{x,y}$ are suppressed due to $|\delta|\gg t_{0,\rm{so}}$, such that the initial state
is polarized to $\mid \uparrow\rangle$ and points to the North pole on the Bloch sphere~\cite{Sun2018}.
Hence, the value of the spin polarization is close to 1 at $t=0$ (See Fig.~1 and Fig~\ref{FigS9}).
Regarding quenching $h_{{\rm F},x}$ or $h_{{\rm F},y}$, we set the phase $(\varphi_1, \varphi_2)=(0, \pi/2)$ or ($\pi/2, 0$), meanwhile the detuning $\delta$ is a constant by fixing the frequency of $\text{RF}_1$, i.e., $\delta=\delta_0$.
Such settings make the Zeeman term constant $m_{x,y}\gg t_{0,\rm{so}}^{x,y}$, which induces $h_{x}$ or $h_y$ dominates as well as the initial state is prepared to be almost polarized in the $\sigma_x$ ($\sigma_y$) direction~\cite{Yi2019}.
In other words, the initial state points near the equator on the Bloch sphere.
Thus, the value of the spin polarization is close to 0 (see Fig.~\ref{FigS9} and Fig.~\ref{FigS4}).
Before all the quenchings, the modulated RF is turned off.
For all the quenchings, the frequency of $\text{RF}_{2}$ is a constant and the phase $\varphi_{1,2}$ are controlled by the $\text{EOM}_{1,2}$.
Thus, the atoms populate mainly in the lowest trivial band of the initial Hamiltonian.

(2) \emph{Quench to the objective Hamiltonian.} For quenching $h_{{\rm F},z}$, the initial detuning is switched to the final detuning $\delta_0\in [-1,1]E_{\rm{r}}$ within 200ns by $\text{Switch}_1$.
Such final detuning is the same order of magnitude as $t_{0,\rm{so}}$, inducing $h_x$ and $h_y$ take effect.
For quenching $h_{{\rm F},x}$ or $h_{{\rm F},y}$, the phase $(\varphi_1, \varphi_2)=(\pi/2,0)$ or $(0,\pi/2)$ is switched to $(\varphi_1, \varphi_2)=(\pi/2, \pi/2)$ by tuning the $\rm{EOM}_1$ or $\rm{EOM}_2$ within 2$\mu \text{s}$.
And the Zeeman constant is switched suddenly from a large $m_x$ or $m_y$ to $m_x=0$ or $m_y=0$, inducing $h_{x,y,z}$ work.
Subsequently, the detuning $\delta$ is modulated for a certain time $t$, i.e., $\delta=\delta_0+A_{\rm{F}}\sin(\omega t)$.
Hence, the atoms evolve in the objective Hamiltonian.

(3) \emph{The detection of the states.} We take off all the lasers in less than 1$\mu$s. Then the Stern-Gerlach magnetic field is turned on so that the atoms in $\mid\uparrow\rangle$ and $\mid\downarrow\rangle$ are separated.
After the atoms expand freely for 25ms, we take photos of atoms to obtain the distribution in $\mid\uparrow\rangle$ and $\mid\downarrow\rangle$ in momentum space, $N_{\uparrow}(\bm{k},t)$ and $N_{\downarrow}(\bm{k},t)$.
Finally, after mapping $N_{\uparrow}(\bm{k},t)$ and $N_{\downarrow}(\bm{k},t)$ to the quasi-momentum space~\cite{Sun2018}, atomic number $N_{\uparrow}(\bm{q},t)$ and $N_{\downarrow}(\bm{q},t)$ are obtained.

\subsection{Band inversion surfaces.}
Band inversion surfaces (BISs) refer to the quasimomenta where the two spin bands are inverted, and can be dynamically identified by resonant spin-flipping oscillations~\cite{Zhang2018}.
For our periodically driven QAH model, the BISs are all 1D rings that surround the $\Gamma$ or $\rm{M}$ point.
We have the following two properties about BISs to distinguish topological phases of the present system:
(P1) The two types of BISs ($0$- or $\pi$-BIS) always appear alternatively along the diagonal direction from $\Gamma$ to $\rm{M}$. For example, in the case of $\mathcal{N}_{\rm{BIS}}=4$ in Fig.4 of main text, the rings from the innermost to outermost are $0$-BIS, $\pi$-BIS, $0$-BIS and $\pi$-BIS, respectively.
(P2) Topological invariant $\nu_{0,\pi}^j$ equals $+1$ (or $-1$) for a BIS surrounding the $\Gamma$ (or $\rm{M}$) point.

\begin{figure}
  \includegraphics[width=1\linewidth]{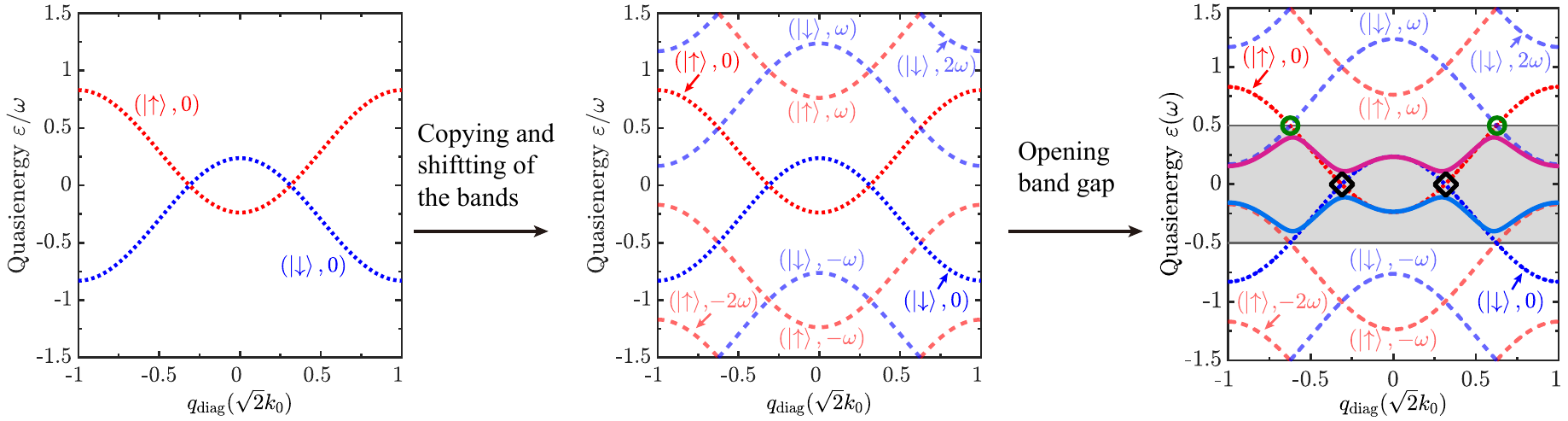}
 \caption{The formation of Floquet bands.
The dotted curves are the static $|\!\uparrow,0\rangle$ and $|\!\downarrow,0\rangle$ bands, which are copied and shifted by the periodic driving (dashes curves). The gap-opening at all band crossings renders the Floquet bands  (solid curves).
  }
  \label{FloquetBandFormation}
\end{figure}

To demonstrate the two conclusions, we write Eq.~(1) as $\mathcal{H}(t)=\mathcal{H}_0+A_{\rm{F}}\sin(\omega t)\sigma_z/2$, and consider the non-driven Hamiltonian $\mathcal{H}_0=\sum_{j=x,y,z}h_j({\bm q})\sigma_j$.
Before demonstrating the two conclusions, let us explain the formation of the Floquet bands.
The spin up and spin down bands are respectively defined as $(\mid\uparrow\rangle,0)$ and $(\mid\downarrow\rangle,0)$, given by $h_z(\bm{q})$ and $-h_z(\bm{q})$, as shown in the first column of Fig.~\ref{FloquetBandFormation}.
In the presence of periodic driving, the periodic driving transfers energies $n\omega$ ($\omega$ is the driving frequency, $n=0,\pm1,\dots$) to the non-driven system,
such that the static spin bands ($(\mid\uparrow\rangle,0)$ and $(\mid\downarrow\rangle,0)$ bands) are copied and shifted in steps of $\omega$.
The copying and shifting lead to new band crossings, corresponding to driving-induced BISs (see the second column of Fig.~\ref{FloquetBandFormation}).
A finite modulation amplitude $A_{\rm{F}}$ and Raman coupling strength $\Omega_0$ then lift the degeneracy and open band gaps, rendering the Floquet bands
shown in the third column of Fig.~\ref{FloquetBandFormation}.
Since the quasi-energy is periodic with the frequency $\omega$, the quasi-energy bands outside the FBZ are redundant $2\pi$-copies.
Similar to the Bloch bands in a lattice system, we only consider the Floquet bands within the FBZ $\varepsilon \in [-\omega/2,+\omega/2]$.

We now demonstrate the two conclusions.
For the first conclusion, we note that the emergence of all BISs comes from the crossings of the spin bands in the Floquet Brillouin zone, determined by $h_z({\bm q})=n\omega/2$ with $n$ being a nonzero integer.
$n$ being even (odd) corresponds to 0-BISs ($\pi$-BISs).
In the absence of periodic driving, the BIS exists at the crossing of $(\mid\uparrow\rangle,0)$ and $(\mid\downarrow\rangle,0)$ bands, where $h_z(\bm q)=0$.
Between two adjacent $\pi$-BISs where $h_z({\bm q})=(2m-1)\omega/2$ and $(2m+1)\omega/2$, there must exist one $0$-BIS that corresponds to $h_z({\bm q})=m\omega$ ($m$ is a integer.), which is ensured by the monotonicity of the spin band $h_z(|{\bm q}|)$.
(See Fig.~1(b) and Fig.~2 in the main text.)

\begin{figure}
  \centering
  \includegraphics[width=0.6\linewidth]{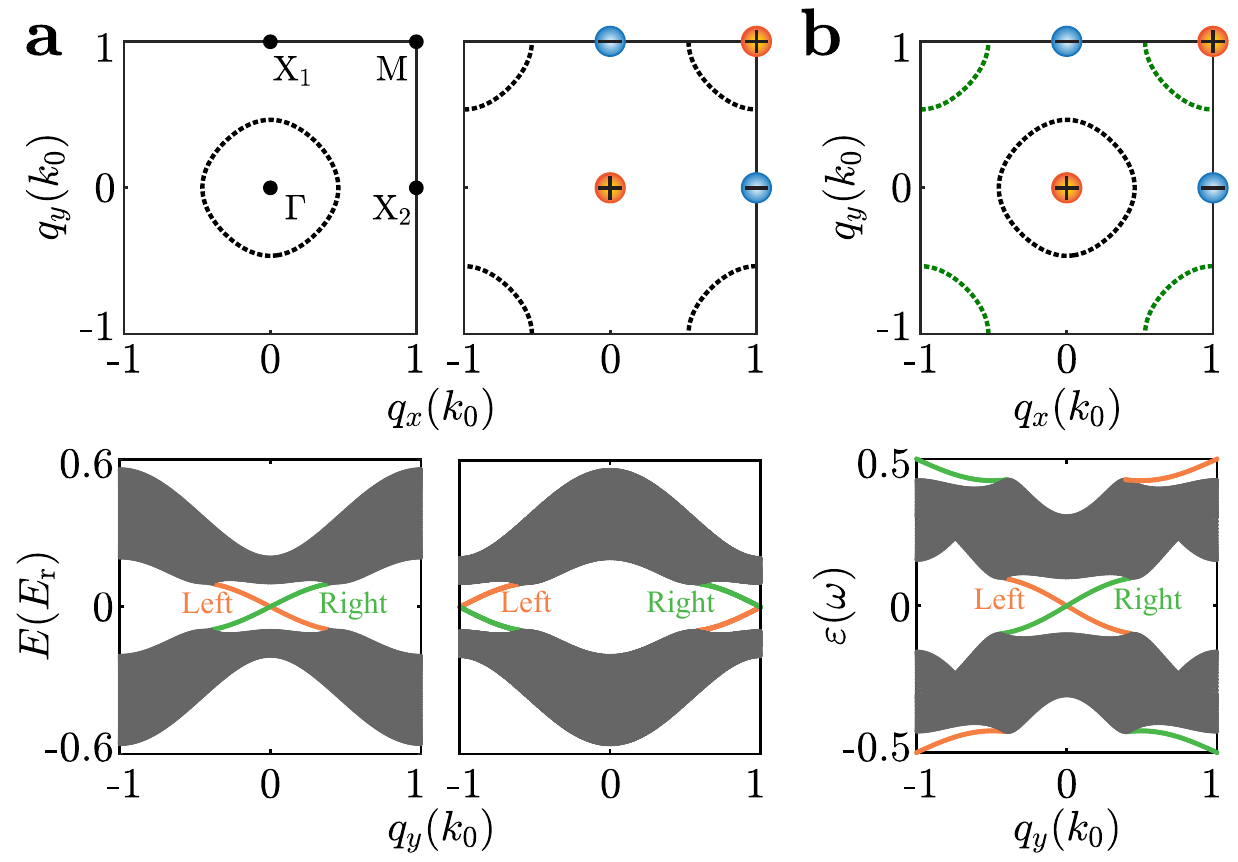}
  \caption{Correspondence between BISs and edge states.
    BISs (upper) and energy spectrum under open boundary conditions (lower) for $\textbf{a}.$ $ \delta_0=0.4E_{\text{r}}$ (left) and $\delta_0=-0.4E_{\text{r}}$ (right) without driving; and $\textbf{b}.$ $\delta_0=0.4E_{\text{r}},T=400\mu\text{s}$.
  }\label{FigS2}
\end{figure}

The second conclusion can be obtained by the aid of topological charges.
For the static Hamiltonian $\mathcal{H}_0$, the topological charges are located at ${\bm q}_{\rm c}$ with $h_x({\bm q}_{\rm c})=h_y({\bm q}_{\rm c})=0$, and the topological invariant defined on a BIS reflects the total charges enclosed~\cite{Zhang2018,Yi2019}.
For the non-driven QAH model, The topological charges are located at four highly symmetric momentum points: $\Gamma$ (charge value ${\cal C}_\Gamma=+1$), $\rm{M}$ (${\cal C}_{\rm M}=+1$) and ${\rm X}_{1,2}$ (${\cal C}_{{\rm X}_{1,2}}=-1$)~\cite{Zhang2018,Yi2019}, as shown in Fig.~\ref{FigS2}\textbf{a}.
When the periodic driving is added to the $h_z$ term of $\mathcal{H}_0$, it does not change the locations and values of the four topological charges (Fig.~\ref{FigS2}\textbf{b}).
Hence, as in the non-driven case, the BIS surrounding the $\Gamma$ point has the topological invariant $\nu_{0,\pi}^j={\cal C}_{\Gamma}=+1$, while
the BIS circling the $\rm{M}$ point has $\nu_{0,\pi}^j={\cal C}_{\Gamma}+{\cal C}_{{\rm X}_1}+{\cal C}_{{\rm X}_2}=-1$.

\subsection{BIS-boundary correspondence.}

Here we demonstrate the property (P3): one-to-one correspondence between BISs and edge states for the 2D periodically driven QAH model under open boundary conditions (OBCs).
The proof consists of two steps: (i) This one-to-one correspondence holds for the non-driven QAH system (Fig. \ref{FigS2}${\textbf a}$). (ii) For the driven model, the Floquet band structure can be regarded as a combination of several static bands with different parameters (Fig.~\ref{FigS2}${\textbf b}$).

The demonstration of Step (i) is straightforward. According to Refs.~\cite{Zhang2018}, the so-called bulk-surface duality reveals a correspondence between the bulk topology and the winding defined on BISs. For the 2D QAH model,
the Chern number ${\cal C}=\pm 1$ is characterised by the topological invariant $\nu=\pm1$
defined on the BIS surrounding the $\Gamma$ (for ${\cal C}=+1$) or $\rm{M}$ (${\cal C}=-1$) point.
Besides, the bulk-edge correspondence obviously holds. We then obtain a one-to-one correspondence between the topological invariant on BIS and the chirality of chiral edge states.

For Step (ii), we examine an approximate expression for the effective Hamiltonian by applying the Floquet-Magnus expansion.
Consider a BIS formed by two shifted bands $(\left| \downarrow \right\rangle,m_1\omega)$ and $(\left| \uparrow \right\rangle,m_2\omega)$ with $\zeta\equiv m_1-m_2$.
We apply a rotation ${\cal O}(t)=\exp(i\zeta\omega t\sigma_z/2)$ to the time-dependent Hamiltonian ${\cal H}(t)$, and write ${\cal H}_{\rm rot}(t)={\cal O}(t){\cal H}(t){\cal O}^\dagger(t)=\sum_{-\infty}^{\infty} H_n e^{in\omega t}$.
According to the Floquet-Magnus expansion~\cite{Eckardt2017}, the effective Hamiltonian near this BIS can be obtained by
${\cal H}_{\rm F}=\sum_{n\neq0}(H_nH_{-n}+[H_0,H_n])/n\omega$, which yields $h_{{\rm F},x/y}\approx (-1)^mJ_m\left(\frac{4A_{\rm{F}}}{\omega}\right)h_{x/y}\propto h_{x/y}$ with the Bessel function $J_m(z)$ and $h_{{\rm F},z}\approx h_z-\zeta\omega/2$ for the driven QAH model.
Hence, the local band structure near a driving-induced BIS can be approximated as the static one with a shifted Zeeman constant $m_{\rm eff}=m_z-\zeta\omega/2$.

With the two results (i) and (ii), we finally come to the conclusion that for our periodically driven QAH model with OBCs, a 0-BIS (or $\pi$-BIS) indicates a chiral edge state within the 0-gap  (or $\pi$-gap).
For example, for the high-Chern-number phase in Fig.~3, the 0-BIS around the $\Gamma$ point corresponds to a chiral edge state in the 0-gap near $q_y=0$, while the $\pi$-BIS around the $\rm M$ point corresponds to a chiral edge state in the $\pi$-gap near $\left|q_y\right|=k_0$.
The fact that the BIS surrounds the $\Gamma$ or $\rm M$ point determines the chirality of the corresponding edge state.

\subsection{Robustness of counterpropagating edge states.}
We shall first consider the ideal tight-binding QAH model
and numerically verify that counterpropagating edge states are robust against smooth impurities.
We then argue that the conclusion also holds for our cold atom realisation beyond the tight-binding limit.

For the periodically driven QAH model, the predicted counterpropagating edge states are protected by both the particle-hole symmetry (PHS) and large-momentum separation.
The PHS ${\cal P}={\cal K}P$, where ${\cal K}$ is the complex conjugation operator and $P$ is unitary, acting on the Hamiltonian gives
$P{\cal H}({\bm q},t)P^{-1}=-{\cal H}^*(-{\bm q},t)$~\cite{Roy2017,Yao2017}, which leads to $\varepsilon({\bm q})=-\varepsilon(-{\bm q})$.
Thus, the zero-energy edge states can only appear at $q_y=0$ or $\pi$ under an OBC in $\hat x$-direction.
The driven QAH model has a PHS $P=\sigma_x$, and the edge states with opposite chirality are located around $q_y=0$ and $\pi$, respectively.
Similar to the quantum valley-Hall effect (see, e.g., Ref.~\cite{Ren2016} and references therein), such a large momentum separation protects
the counterpropagating edge states from long-range disorders.

\begin{figure}
  \centering
  \includegraphics[width=0.5\linewidth]{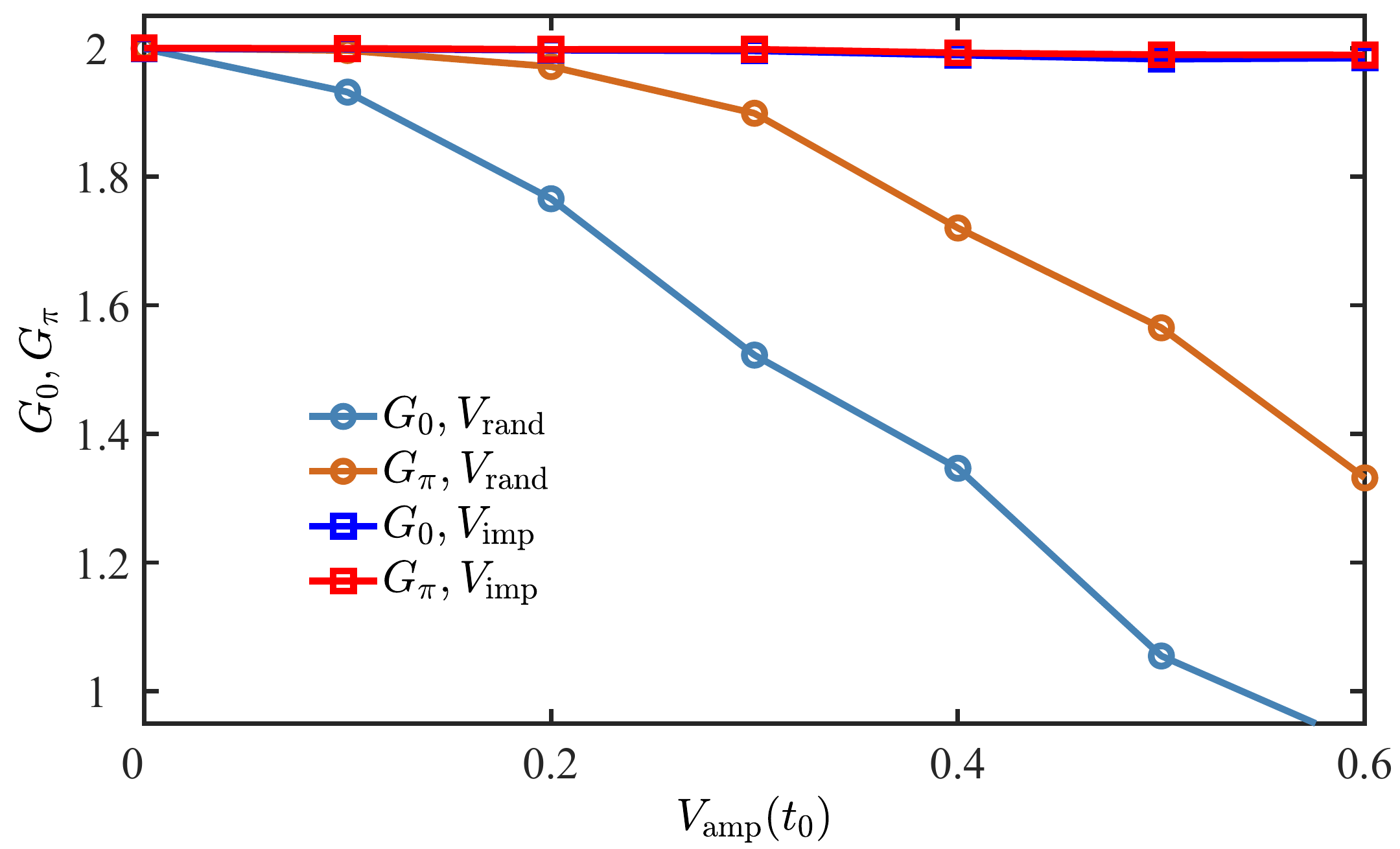}\\
  \caption{Quasienergy-dependent conductance $G_{0,\pi}$ in the presence of two kinds of disorders.
    Here we take the Hamiltonian as the QAH tight-binding model Eq.~(1) with $\delta_0=6t_0$, $t_{\rm so}=t_0$, and
    the driving frequency $\omega=4t_0$. The lattice size is set as $40\times40$.
    Each point is averaged over 50 disorder configurations.
  }\label{FigS3}
\end{figure}

To examine the disorder effects, we employ scattering matrix invariants to characterise the topological phase~\cite{Fulga2016,Umer2020}.
The quasienergy-dependent Floquet scattering matrix can be expressed as
\begin{align*}
S(\varepsilon)={\sf P}\left[e^{-i\varepsilon}{\bm 1}-U(1-{\sf P}^{\rm T}{\sf P})\right]^{-1}U{\sf P}^{\rm T},
\end{align*}
where ${\sf P}$ is the projector matrix (see Ref.~\cite{Fulga2016} for the definition), the superscript ${\rm T}$ denotes the matrix transpose, and $U$ is the
time-evolution operator under open boundary conditions.
The scattering matrix takes the form
\begin{align*}
S(\varepsilon) =
\left( \begin{array}{cc}
r & t \\
t' & r'\\
\end{array} \right).
\end{align*}
From this one can compute the conductance $G_{0,\pi}={\rm Tr}(tt^\dagger)$, which characterises
the number of edge states within the corresponding quasienergy gap.
We consider two kinds of PHS-preserving disorders applied to the Zeeman term, i.e., $m_z\to m_z+V_{\rm disorder}$. The first one is the random on-site disorder
\begin{align*}
V_{\rm rand}({\bm r})=\sum_j V_j\delta({\bm r}-{\bm r}_j),
\end{align*}
and the second one represents a long-range potential induced by smooth impurities
\begin{align*}
V_{\rm imp}({\bm r})=\sum_l^{N_{\rm imp}}\frac{V_l}{\sqrt{({\bm r}-{\bm r}_l)^2+d^2}},
\end{align*}
where $N_{\rm imp}$ denotes the number of the randomly distributed impurities
and $d$ controls the potential range.
Here we set $d=1$ and take $V_{\rm rand},V_{\rm imp}\in[-V_{\rm amp},V_{\rm amp}]$, with $V_{\rm amp}$ denoting the strength of disorders.
The numerical results are shown in Fig.~\ref{FigS3} for disorder effects in the phase with ${\cal N}_{\rm BIS}=4$.
Without disorders, two $0$-BIS (or $\pi$-BISs) with opposite topological invariants $\pm1$ ensure a pair of counterpropagating edge states within the 0-gap (or $\pi$-gap),
due to the correspondence between BISs and edge states.
In the presence of on-site disorders $V_{\rm rand}({\bm r})$, our numerical results show that $G_{0,\pi}$ deviate from their quantized values 2 as the disorder strength $V_{\rm amp}$ increases,
indicating that the edge states with opposite chirality are hybridized with each other by disorders.
In comparison, the computed values of $G_{0,\pi}$ are almost unaffected by $V_{\rm imp}({\bm r})$,
which verifies that the counterpropagating edge states are immune against long-range disorders.

The above conclusion on the robustness of counterpropagating edge states is not restricted to the ideal QAH model.
For our realistic system beyond the tight-binding limit, the PHS is broken by the next-nearest-neighbor hopping.
However, the correspondence between BISs and edge states still holds, and each gapless edge state is protected by a valley-like local topological structure formed by a BIS~\cite{zhang2021}.
Hence, the large momentum separation between two ``valleys'' also protects the counterpropagating edge states against smooth impurities.

\begin{figure}
  \includegraphics[width=0.9\linewidth]{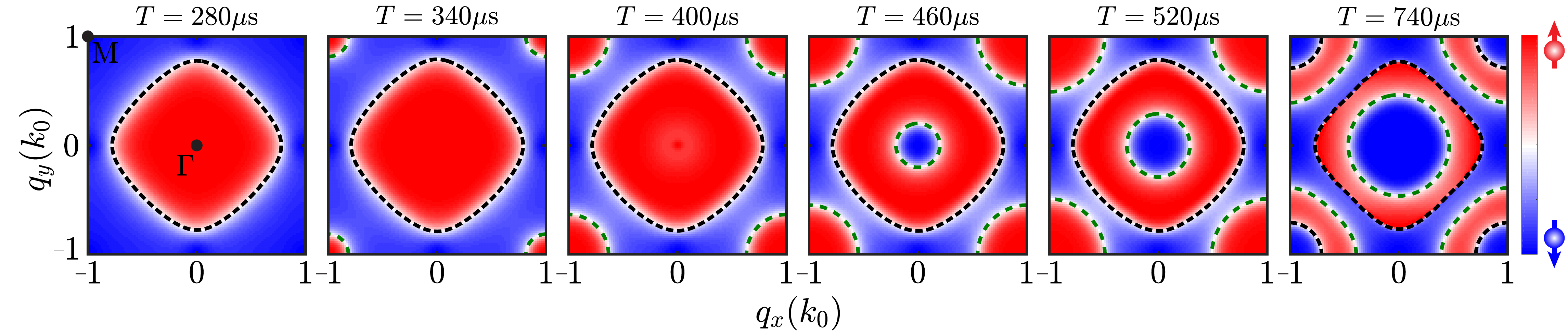}
 \caption{The spin textures $\left\langle \sigma_z(\bm{q}) \right\rangle$ for different driving period $T$.
  $\Gamma$ and M are high symmetric momenta.
  The black and green dashed curves denote 0-BIS and $\pi$-BIS, respectively.
  The red (blue) colour in the spin textures denotes spin up (down).
  The parameters $(V_0,~\Omega_0,~\delta_0,A_{\text{F}})=(4.0,~1.0,~0.1,~0.8)E_{\text{r}}$.
  }
  \label{FigSpinTextureStable}
\end{figure}

\subsection{
Obtainment of the time-averaged spin textures
}

To characterise the topology of driven QAH model, the critical step is to obtain the time-averaged spin textures $\overline{\left\langle\sigma_z(\bm{q},t)\right\rangle}_{x,y,z}$.
To this end, we fit the time-evolved spin textures $\left\langle\sigma_z(\bm{q},t)\right\rangle_{x,y,z}$ at each quasi-momentum point $\bm{q}$ by the combination function (also see Ref.\cite{Yi2019})
\begin{equation}
\begin{aligned}
\label{eqSdecay}
P_{z}(\bm{q},t)=\sum_{i=1}^{2}A_i(\bm{q})\cos(2\pi \nu_i(\bm{q})t+\varphi_i)e^{-\frac{t}{\tau_1}}+B(\bm{q})e^{-\frac{t}{\tau_2}}+D(\bm{q}),
\end{aligned}
\end{equation}
where the $A_i(\bm{q})$ term represents the damped oscillations with characteristic frequency $\nu_i$ and damping time $\tau_1$, the $B(\bm{q})$ term represents the pure decay with characteristic time $\tau_2$, and the $D(\bm{q})$ term is the offset. The damping and the decay are induced by the noise of magnetic field and the interaction between the atoms. Then we remove these relaxation effects and calculate the average of spin polarization by

\begin{figure}
  \includegraphics[width=0.6\linewidth]{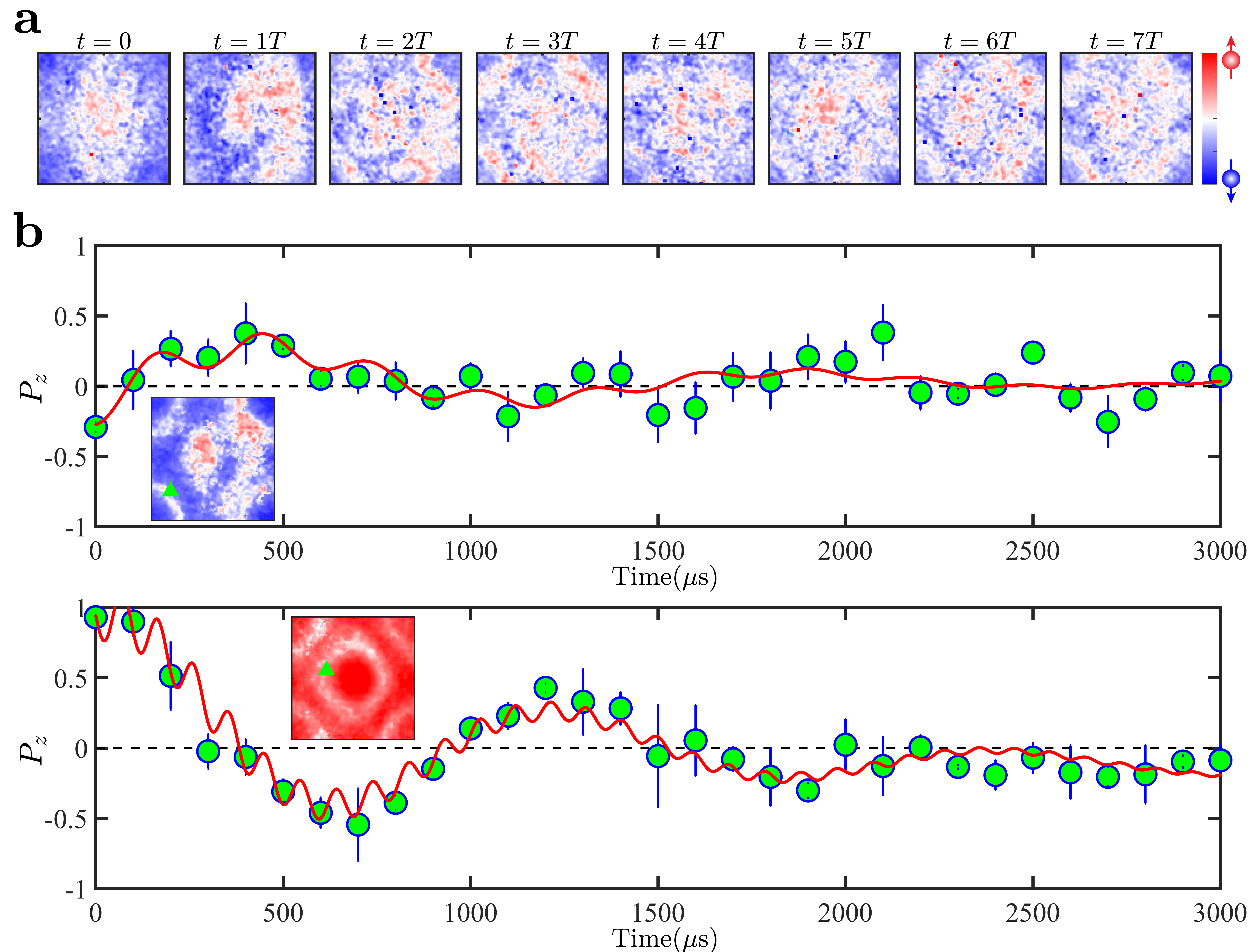}
  \caption{\textbf{a.} Time-evolved spin textures for quenching $h_{{\rm F},x}$.
  \textbf{b.} Time-evolved spin polarisation for quenching $h_{{\rm F},x}$ (upper) and $h_{{\rm F},z}$ (lower).
  The green circles with error bars are experimental data. The red curves are the fitting curves.
  The insets are the time-averaged spin textures $\overline{\left\langle\sigma_z(\bm{q},t)\right\rangle}_x$ ($\overline{\left\langle\sigma_z(\bm{q},t)\right\rangle}_z$), on which the quasi-momentum points $\bm{q}$ are marked by green triangles.
  Parameters are $(V_0,~\Omega_0,~\delta_0,~A_{\rm F})=(4.0,~1.0,~0.4,~0.8)E_{\text{r}}$ and $T=400\mu \text s$.
  }
  \label{FigS4}
\end{figure}

\begin{equation}
\begin{aligned}
\label{eqSideal}
\overline{\langle \sigma_{z}(\bm{q})\rangle}_{x,y,z}=
\frac{1}{NT}\sum_{t=0}^{NT} \left(\sum_{i=1}^{2}A_i(\bm{q})\cos(2\pi \nu_i(\bm{q})t+\varphi_i)\right)+B(\bm{q})+D(\bm{q})
\end{aligned}
\end{equation}

where $T$ and $N$ are the driving period and the integer number, respectively.
For example, for quenching $h_{{\rm F},x}$ (time-evolved spin textures shown in Fig.~\ref{FigS4}$\textbf{a}$), the time-evolved spin polarisation at $\bm{q}=(-0.70,0.51)k_0$ is shown in Fig.~\ref{FigS4}$\textbf{b}$ (upper).
According to the fitting parameters and setting $NT=107\times 280\mu s$, we have $\overline{\langle \sigma_{z}(\bm{q}=(-0.70,0.51)k_0)\rangle}_{x}\approx 0.16$.
Similarly, for quenching $h_{{\rm F},z}$, we have $\overline{\langle \sigma_{z}(\bm{q}=(-0.43,-0.19)k_0)\rangle}_{z}\approx -0.12$ (Fig.~\ref{FigS4}$\textbf{b}$ (lower)).

\subsection{The sign of $h_{\text{F},z}$}

In Fig.~3$\textbf{a}$ of the main text, the two BISs divide the first Brillouin zone into three regions.
The sign of $h_{\text{F},z}$ in each region can be determined by the formation of the Floquet bands.
Consider the static Hamiltonian $\mathcal{H}_0=\sum_{j=x,y,z}h_j({\bm q})\sigma_j$, of which the $\mid\uparrow\rangle$ and $\mid\downarrow\rangle$ bands cross each other at the momenta where $h_z(\bm q)=0$.
when periodic driving is added to $\mathcal{H}_0$, the spin bands are copied and shifted by $n\omega$ ($n$ is an integer), forming band crossings where $h_z({\bm q})=n\omega/2$.
The finite Raman coupling strength $\Omega_0$ and driving amplitude $A_{\rm{F}}$ open gaps at these crossings, so that band inversion surfaces (BISs) emerge, and the spin bands folded in the Floquet Brillouin zone form the Floquet bands $h_{\text{F},z}$ (Fig.~2).
Thus, the sign of $h_{\text{F},z}$ , which is the same as the sign of spin polarisation $-\left\langle \sigma_z \right\rangle$, must be opposite on both sides of the BIS (Fig.~\ref{FigSpinTextureStable}).
Therefore, in Fig.~3$\textbf{a}$ of the main text, the sign of $h_{\text{F},z}$ in the three regions from $\Gamma$ to $\text{M}$ are "-", "+" and "-", respectively.

\subsection{The dynamical field and the topological invariant $\nu_{0,\pi}^j$}
\begin{figure}
  \center
  \includegraphics[width=0.3\linewidth]{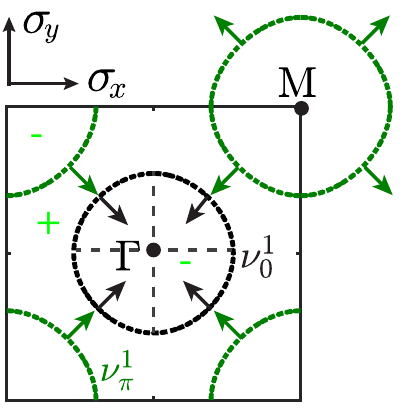}
 \caption{The obtainment of $\nu_{0,\pi}^j$.
  }
  \label{Fig_nux}
\end{figure}

We obtain $\nu^j_{0,\pi}$ from the following steps:

(i) Constructing the dynamical field from time-averaged spin textures.
According to Ref.~[23], the dynamical field $\bm{g}(\bm{q})=(g_y,g_x)$ is defined with two components $g_{x,y}(\bm{q})\equiv-\frac{1}{{\cal N}_{\bm q}}\partial_{q_{\perp}}\overline{\langle\sigma_{z}(\bm{q})\rangle}_{x,y}$, where
$\mathcal{N}_{\bm{q}}$ is a normalization factor and $q_{\perp}$ denotes the momentum perpendicular to the BIS and points from ``$-$'' to ``$+$''.
Here the sign ``$+$''  (``$-$'') denotes the region where $h_{{\rm F},z} >0$ ($h_{{\rm F},z}<0$), which can be judged by the Floquet band structure and the properties of band inversion (see the Supplemental Materials for details).
From the definition of $g_{x,y}(\bm{q})$, one can see that the direction of the dynamical field is determined by the difference
of the measured spin polarization between the two regions divided by the BIS.
Due to the minus sign in the definition, we compute the directional derivative across the BIS by subtracting the measured spin polarization of the ``$+$'' region from the measured spin polarization of the ``$-$'' region.
Take Fig.~3(a) in the main text as an example. The region B is the ``$+$'' region and A is the ``$-$'' region. The polarization difference across the BIS is thus $P(A)-P(B)<0$, where $P(A)$ [$P(B)$] denotes the measured spin polarization in region $A$ ($B$). Hence, we have $g_{x}(\bm{q})<0$ near the region $A$ and $B$, which means that the direction of the $x$-component of the dynamical field should be in the opposite direction of $\sigma_x$-axis.
Like this, the components of the dynamical field $g_{x,y}$ are obtained and drawn in the first two columns of Fig. 3 in the main text.
Thus, superimposing the two components $g_{x,y}$, the combined dynamical field $\bm{g}$ is obtained, i.e., $\bm{g}(\bm{q})=g_x(\bm{q})\hat{e}_x+g_y(\bm{q})\hat{e}_y$, as shown in the three column of Fig. 3 in the main text.

(ii) Determining the topological invariant from the dynamical field.
According to Ref.~\cite{Zhang2020}, the topological invariant associated with the $j$-th BIS is defined as $\nu^j_{0,\pi}=\int_{\rm{BIS}}\bm{g}(\bm{q}){\rm d}\bm{g}(\bm{q})/2\pi$,
where ${\bf g}{\rm d}{\bf g}\equiv g_{y}{\rm d}g_{x}-g_{x}{\rm d}g_{y}$ and ``${\rm d}$'' denotes the exterior derivative,
which characterizes the winding number of the dynamical field $\bm{g}(\bm{q})$ on BISs.
The expression also tells that the winding orientation of the dynamical field along the BIS determines whether the associated topological number is positive or negative.
As shown in Fig.~\ref{Fig_nux}, we need to not only count how many times these arrows rotate around along each BIS,
but also examine the direction of their rotation (clockwise or anticlockwise).
Note that the winding orientation should be identified with respect to the same reference point. Here we take the $\Gamma$ point as the reference point, and first consider the BIS surrounding $\Gamma$  (black circle). When tracing out the black circle anticlockwise around $\Gamma$, one can find that the dynamical field on the BIS winds once also in an anticlockwise direction, which indicates that the associated topological number should be $\nu^1_{0}=+1$.
In contrast, for the BIS surrounding the $M$ point (green circle), the situation is reversed.
When we trace out the green circle anticlockwise with respect to $\Gamma$, we actually make a circle that travels {\it clockwise} around M (the upper right in Fig.~\ref{Fig_nux}).
Therefore, the winding orientation of the dynamical field also becomes clockwise, giving the topological number  $\nu^1_{\pi}=-1$.

Based on above descriptions, one sees that the experimental data are used to determine both the sign of $\nu^j_{0,\pi}$ and also its magnitude.
Since it characterizes how many times the dynamical field winds along the BIS (Here, selecting the same momentum point for the start and end point on the dynamical field to obtain $\nu^j_{0,\pi}$.), the values of $\nu^j_{0,\pi}$ must be quantized without error bars.

\subsection{Four BISs in momentum space}
\begin{figure}
  \center
  \includegraphics[width=0.5\linewidth]{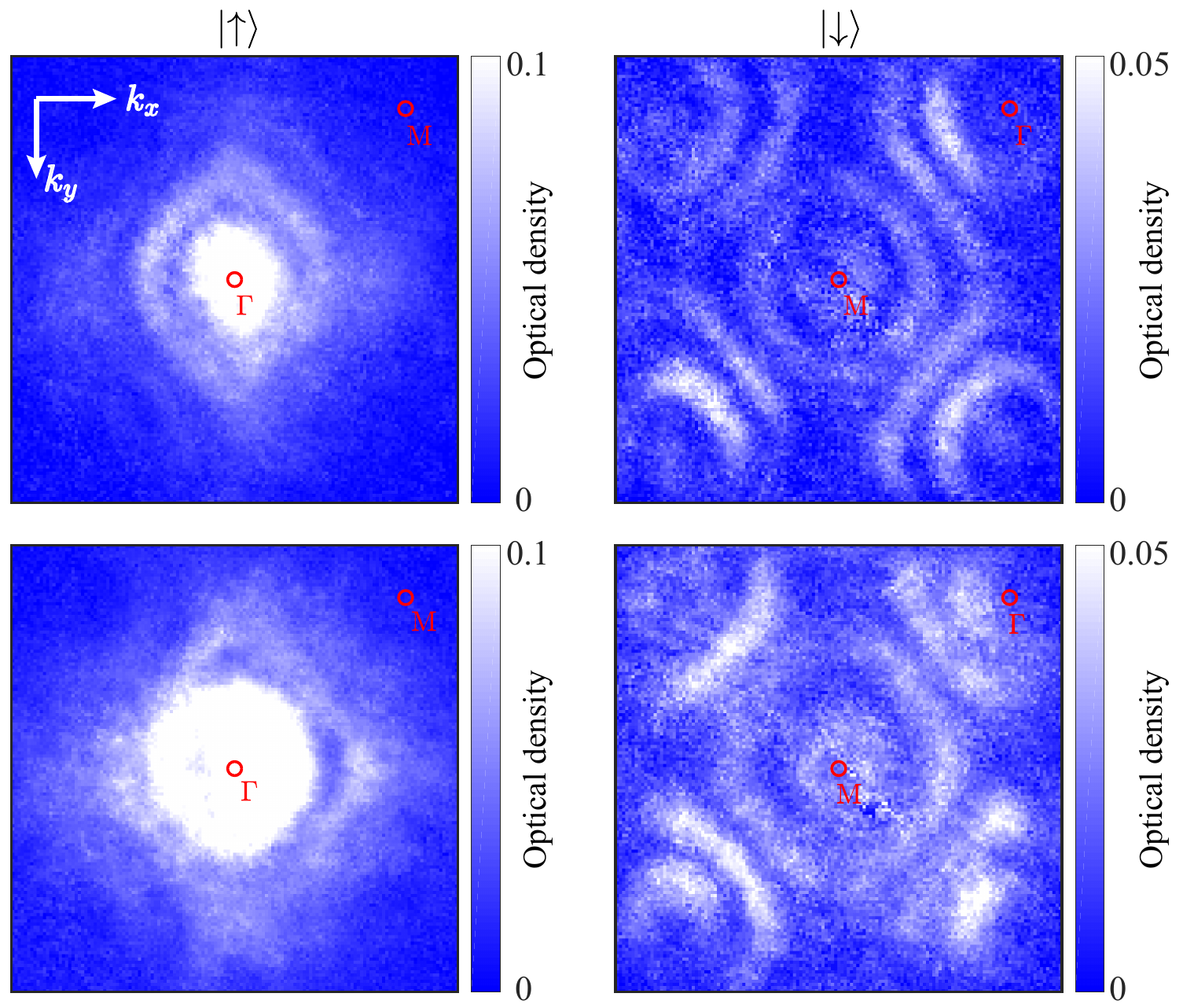}
 \caption{$N_{\uparrow}(\bm{k},t)$ and $N_{\downarrow}(\bm{k},t)$ in momentum space.
 The upper (lower) row corresponds to four BISs data of Fig. 2 (Fig. 4) before mapping to quasi-momentum space with $t=2T$.
 The parameters of upper (lower) row is $V_0=4.0E_{\rm{r}},\Omega_0=1.0E_{\rm{r}},\delta_0=0.1E_{\rm{r}},T=740\mu s,A_{\rm{F}}=1.2E_{\rm{r}}$ ($V_0=4.0E_{\rm{r}},\Omega_0=1.0E_{\rm{r}},\delta_0=-0.2E_{\rm{r}},T=600\mu s,A_{\rm{F}}=1.2E_{\rm{r}}$).
  }
  \label{Fig4ring}
\end{figure}

For quenching $h_{\text{F},z}$, the atoms are prepared in the initial polarized state $\mid \uparrow \rangle$.
After quenching $h_{\text{F},z}$, the spin flips appear and the atoms flip to $\mid\downarrow\rangle$.
Those atoms flipping to $\mid\downarrow\rangle$ form ring structures surrounding $\Gamma$ or M, which is the signature of the BISs.
The experimental data with four BISs in momentum space are shown in Fig.~\ref{Fig4ring}.
From the momentum distribution of $\mid\downarrow\rangle$ $N_{\downarrow}(\bm{k},t)$, four rings are clearly observed.

\subsection{Determination of the topological phase boundary}
\begin{figure}
  \includegraphics[width=0.5\linewidth]{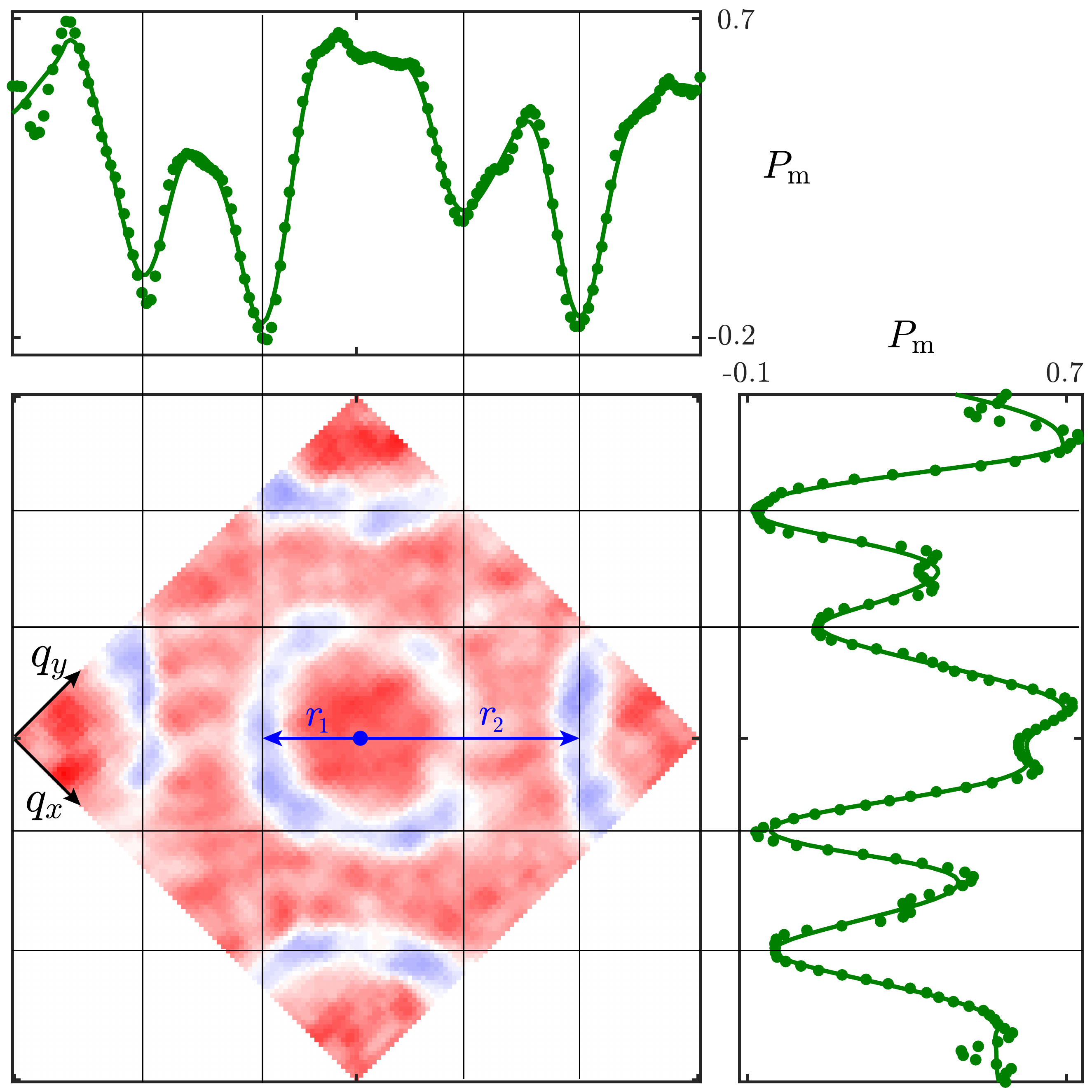}
 \caption{The extraction of the size $r$.
 The spin texture is obtained via rotating the original spin texture by $45^{\circ}$.
 The right and top subfigures are obtained by extracting the spin polarization along the diagonal and anti-diagonal direction of quasi-momentum $q_{\rm{diag}}$, respectively.
 The green dots are experimental data and the solid curves are the fitting by the numerical smoothing.
 The sizes $r_{1,2}$ are obtained by measuring the distance from $\Gamma$ point to the local minimal value of $P_{\rm{m}}$.
 The parameters: $V_0=4.0E_{\rm{r}},\Omega_0=1.0E_{\rm{r}},\delta_0=0.4E_{\rm{r}},T=400\mu s,A_{\rm{F}}=0.8E_{\rm{r}},t=3T$.
  }
  \label{FigExtractRing}
\end{figure}

According to the theoretical analysis, BIS is the essence of the topological phases in the driven QAH model \cite{Zhang2020}.
The BISs vanish for the trivial phases, while they exist for the topological phases.
Moreover, the size of BISs change continuously when one of the parameters $V_0$, $\Omega_0$, $T$ and $\delta_0$ is tuned.
Thus, via tuning $T$ and $\delta_0$ as well as fixing $V_0$ and $\Omega_0$, the topological phase boundaries in Fig.~4(a) in the main text can be determined where the corresponding BIS just appears.
Phase boundaries with black straight lines in Fig.~4(a) are identified by measuring the size of BISs as a function of the detuning $\delta_0$, while other phase boundaries are extracted by measuring the size of BISs as a function of the period $T$.
Here, the size of the ring $r$ is defined as the distance between $\Gamma$ point and the ring.

\begin{figure}
  \includegraphics[width=1\linewidth]{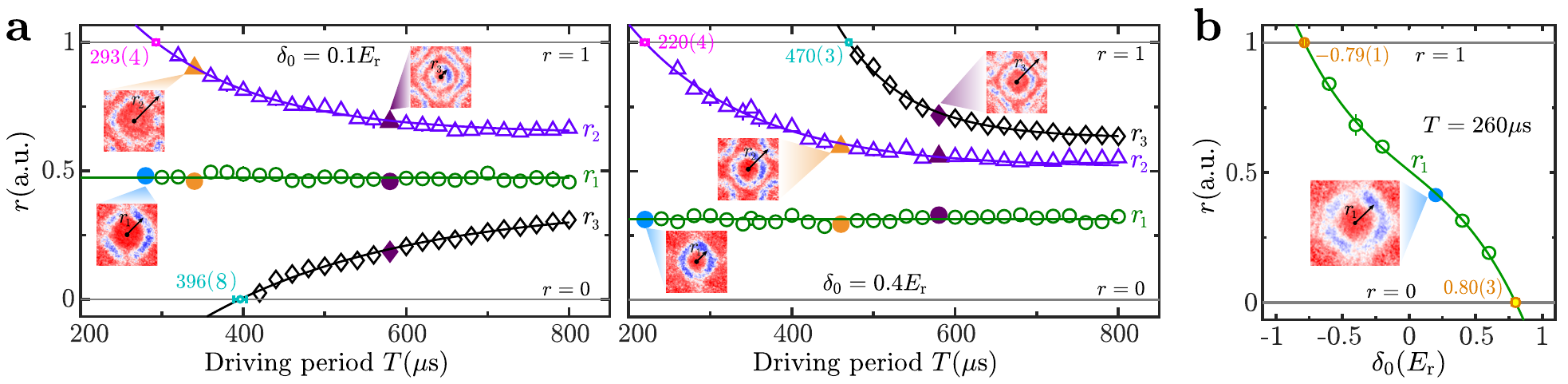}
  \caption{Topological phase boundary.
  $\textbf{a.}$ Topological phase boundaries for $(V_0, \Omega_0)=(4.0, 1.0)E_\text{r}$, $\delta_0=0.1E_\text{r}$ (left) and $\delta_0=0.4E_\text{r}$ (right). Green circles denote the size of the ring $r_1$ corresponding to the BIS of static QAH bands; purple triangles and black diamonds represent the sizes of rings $r_2$ and $r_3$ induced by periodic driving, which are fitted exponentially. The fitting curves intersect the straight lines $(r=0)$ or $(r=1)$, which gives the location of phase boundaries (small blocks). Insets are the representative spin textures with different number of rings.
  $\textbf{b.}$ Topological phase boundaries for $(V_0, \Omega_0)=(4.0, 1.0)E_\text{r}$ and $T=260\mu\text{s}$.
  }
  \label{FigS6}
\end{figure}

The size $r$ is extracted by locating the local minimal of the spin polarization along the diagonal and anti-diagonal direction of the quasi-momentum $q_{\rm{diag}}$.
The procedure of the extraction is the following:
I. Extracting the spin polarization along the diagonal and anti-diagonal direction of the quasi-momentum $q_{\rm{diag}}$, as shown in Fig.~\ref{FigExtractRing}.
II. Obtaining four $r_1$ and four $r_2$ by measuring the distance between the quasi-momentum point (where the spin polarization $P_{\rm{m}}$ is the local minimal value) and $\Gamma$ point.
III. Repeating the steps I and II to obtain the expected values and the statistic errors of the sizes for multiple spin textures with the same parameters.
IV. Adjusting the modulation period $T$ or the detuning $\delta_0$, the sizes are extracted, as shown in Fig.~\ref{FigS6}.

As the driving period $T$ increases, there emerge multiple rings in the time-evolved spin textures (Fig.~\ref{FigS6}$\textbf{a}$).
The first ring corresponds to BIS formed by the spin-orbit coupling of static QAH model, of which the size $r_1$ is constant.
Other rings correspond to the BISs induced by the driving, of which the sizes $r_{2,3}$ vary.
We fit $r_{2,3}$ with an exponential function
\begin{equation}
\begin{aligned}
\label{eqS3}
r=Ae^{-\frac{T}{T_0}}+r_0,
\end{aligned}
\end{equation}
where $r_0$ and $T_0$ are the fitting parameters.
The topological boundary is the intersection between the fitting curve and the straight line $r=0$ or $r=1$.
$r=0$ ($r=1$) represents the smallest (largest) size of the ring where the ring just emerges.
When the detuning $\delta_0=0.1 (0.4) E_{\rm{r}}$, The topological boundaries are $T=(293\pm 4)\mu\text{s}$ ($T=(220\pm 4)\mu\text{s}$) for the number of rings from one to two and $T=(396\pm 8)\mu\text{s}$ ($T=(470\pm 3)\mu\text{s}$) for the number of rings from two to three.

On the other hand, as the detuning $\delta_0$ increases, the size of the ring $r_1$ in the time-evolved spin textures decreases (Fig.~\ref{FigS6}$\textbf{b}$).
To obtain the topological boundaries with black straight lines in Fig.~4$\textbf{a}$, we fit the size $r_1$ by a polynomial function \cite{Sun2018}
\begin{equation}
\begin{aligned}
\label{eqS3}
r=A+B\delta_0+C\delta_0^3,
\end{aligned}
\end{equation}
where $A$, $B$ and $C$ are the fitting parameters.
The topological boundary is also the intersection between the fitting curve and the straight line $r=0$ or $r=1$.
Thus, when $T=260\mu\text{s}$, the topological boundaries are $\delta_0=(0.8\pm 0.03)E_{\rm{r}}$ and $\delta_0=(-0.79\pm 0.01)E_{\rm{r}}$ (Fig. \ref{FigS6}$\textbf{b}$).

\begin{figure}
  \includegraphics[width=0.9\linewidth]{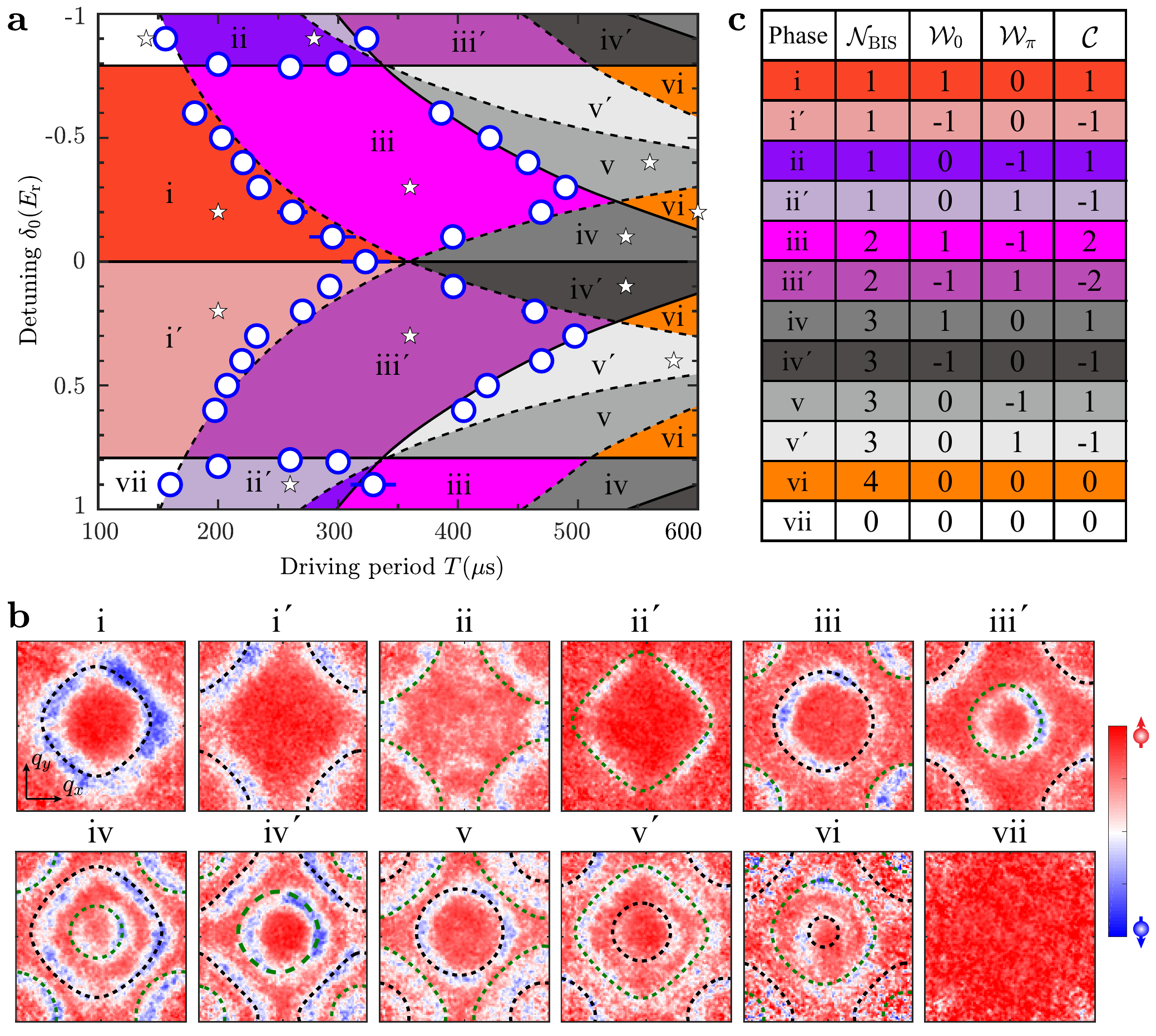}
  \caption{\textbf{a.} Topological phase diagram for $(V_0, \Omega_0)=(4.0, 1.0)E_\text{r}$.
The blue circles with error bars are experimental measurements.
The solid or dashed curves of numerical calculations denote phase boundaries on which a 0-BIS or a $\pi$-BIS appears, respectively.
The Roman numerals {\romannumeral1} to {\romannumeral12} label twelve topological phases in the experimental parameters.
\textbf{b.} The spin textures $\left\langle \sigma_z \right\rangle_z$ for each phase marked by {\text{\ding{73}}} in \textbf{a}.
The dashed curves with different colours distinguish the twelve phases.
$t=(3,4,3,2,3,2,3,2,3,3,5,3)T$ from the phase {\romannumeral1} to {\romannumeral12}.
$A_{\rm{r}}=0.9E_{\rm{r}}$ for the phase {\romannumeral6} and $A_{\rm{r}}=0.8E_{\rm{r}}$ for other phases.
\textbf{c.} Topological invariants of the twelve phases.
  }
  \label{FigS_phase_diagram}
\end{figure}

\subsection{The topological phase diagram}
Based on the method of determined the phase boundaries descried in section \uppercase\expandafter{\romannumeral1}\textbf{.C}, we summarize the topological phase diagram of driven QAH model in Fig.~\ref{FigS_phase_diagram}\textbf{a}, part of which has been shown in Fig.4 in the main text.
We can reach twelve phases by varying driving period $T$ and detuning $\delta_0$.
The typical spin textures $\left\langle \sigma_z \right\rangle_z$ for each phase are shown in Fig.~\ref{FigS_phase_diagram}\textbf{b}.

Further, we demonstrate the topological invariants of the twelve phases in Fig.~\ref{FigS_phase_diagram}\textbf{c}, including the total number of BISs $\mathcal{N}_{\rm BIS}$, and the winding number of 0-gap ($\pi$-gap), i.e., $\mathcal{W}_0$ $(\mathcal{W}_\pi)$. And the Chern number is obtained by $\mathcal{C}=\mathcal{W}_0-\mathcal{W}_{\pi}$.
The phases (i-v) can be transformed into ($\text{i}^{\prime}-\text{v}^{\prime}$) by changing the sign of the winding number $\mathcal{W}_0$ or $\mathcal{W}_{\pi}$.
And the phases in $\delta_0>0$ plane and $\delta_0<0$ plane can be also converted to each other.
For example, the winding number $(\mathcal{W}_{0},\mathcal{W}_{\pi})=(+1,-1)$ of the phase (iii) with $\delta_0<0$ can be transfers into $(\mathcal{W}_{0},\mathcal{W}_{\pi})=(-1,+1)$ of the phase ($\text{iii}^{\prime}$) with $\delta_0>0$.

\begin{figure}
  \center
  \includegraphics[width=0.8\linewidth]{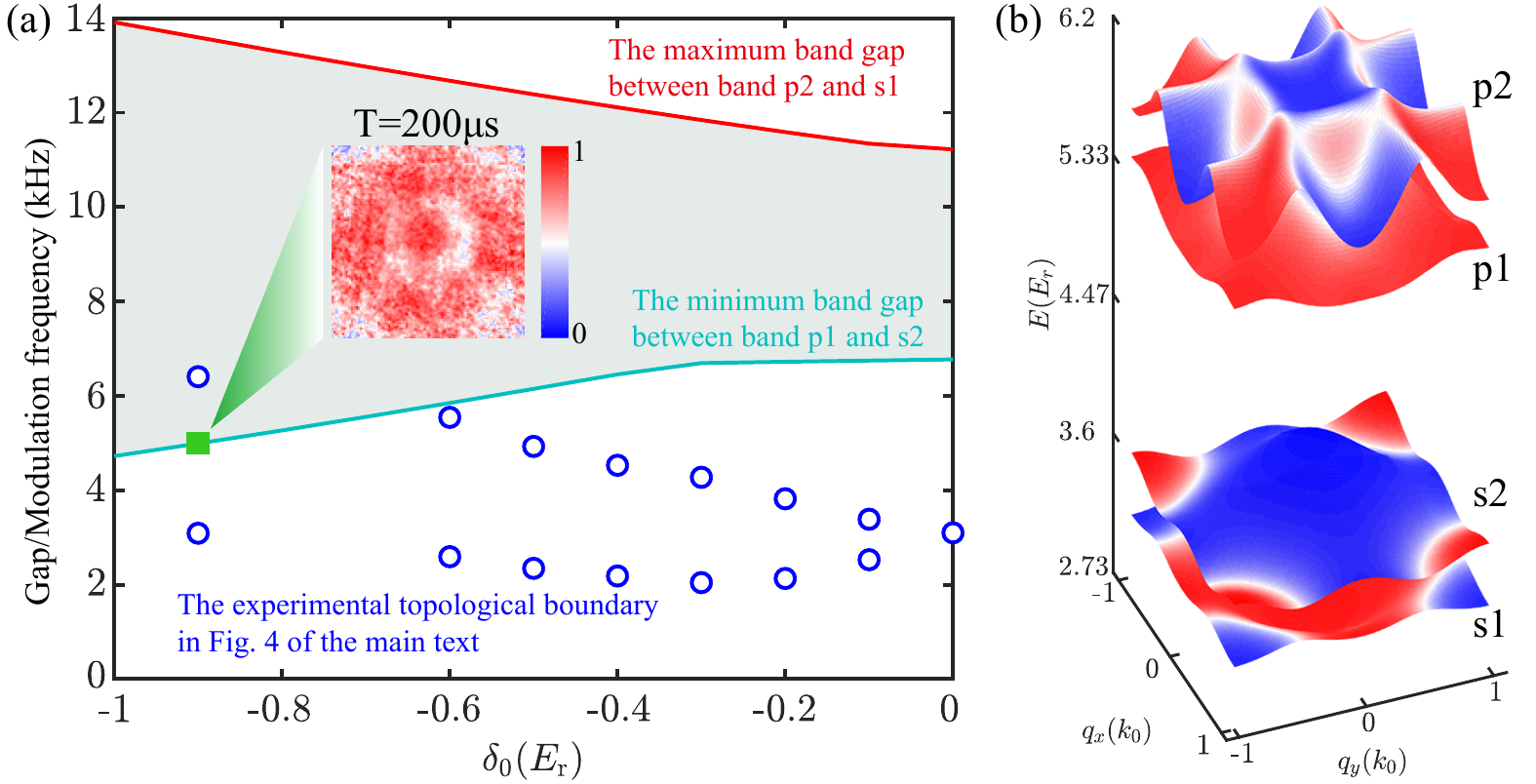}
 \caption{(a) Energy gaps to the first excited bands compared to the modulation frequency.
 The red and green solid curves are respectively the upper and lower bounds of the gaps between the first two excited bands ($p_{1,2}$) and the lowest two bands ($s_{1,2}$).
 The dots represent the parameter settings in our measurents shown in Fig. 4(a) of the main text.
Inset: The measured spin texture with the temperature of 100nK.
Here the parameters are $\delta_0=-0.9E_{\rm{r}}$, the modulation period $T=200\mu s$, and the amplitude $A_{\text{F}}=0.8E_{\rm{r}}$. The measurement is performed at time $t=4T$.
 (b) The calculated band structure of our 2D Raman lattice with two lowest bands $s_{1,2}$ and two excited bands $p_{1,2}$.
  }
  \label{FiggapAndModulationFrequency}
\end{figure}

In addition, the energy gaps to the first excited bands are shown in Fig.~\ref{FiggapAndModulationFrequency}.
The selected experimental parameters are mostly beneath the lower bound (green curve) of the gaps between the first two excited bands ($p_{1,2}$) and the lowest two bands ($s_{1,2}$), except one case with $|\delta_0|=0.9E_{\text{r}}$, which locates slightly above the lower bound.
For $\delta_0=-0.9E_{\text{r}}$, only the BIS surrounding $\Gamma$ point appears.
The higher band effect only appears near the M point, which, however, does not affect the extraction of the information of the BIS near the $\Gamma$ point, see the inset in Fig.~\ref{FiggapAndModulationFrequency}(a).
For $\delta_0=0.9E_{\text{r}}$, the case is similar to $\delta_0=-0.9E_{\text{r}}$.
Thus, the full phase diagram was mapped.

\begin{figure}
  \includegraphics[width=0.7\linewidth]{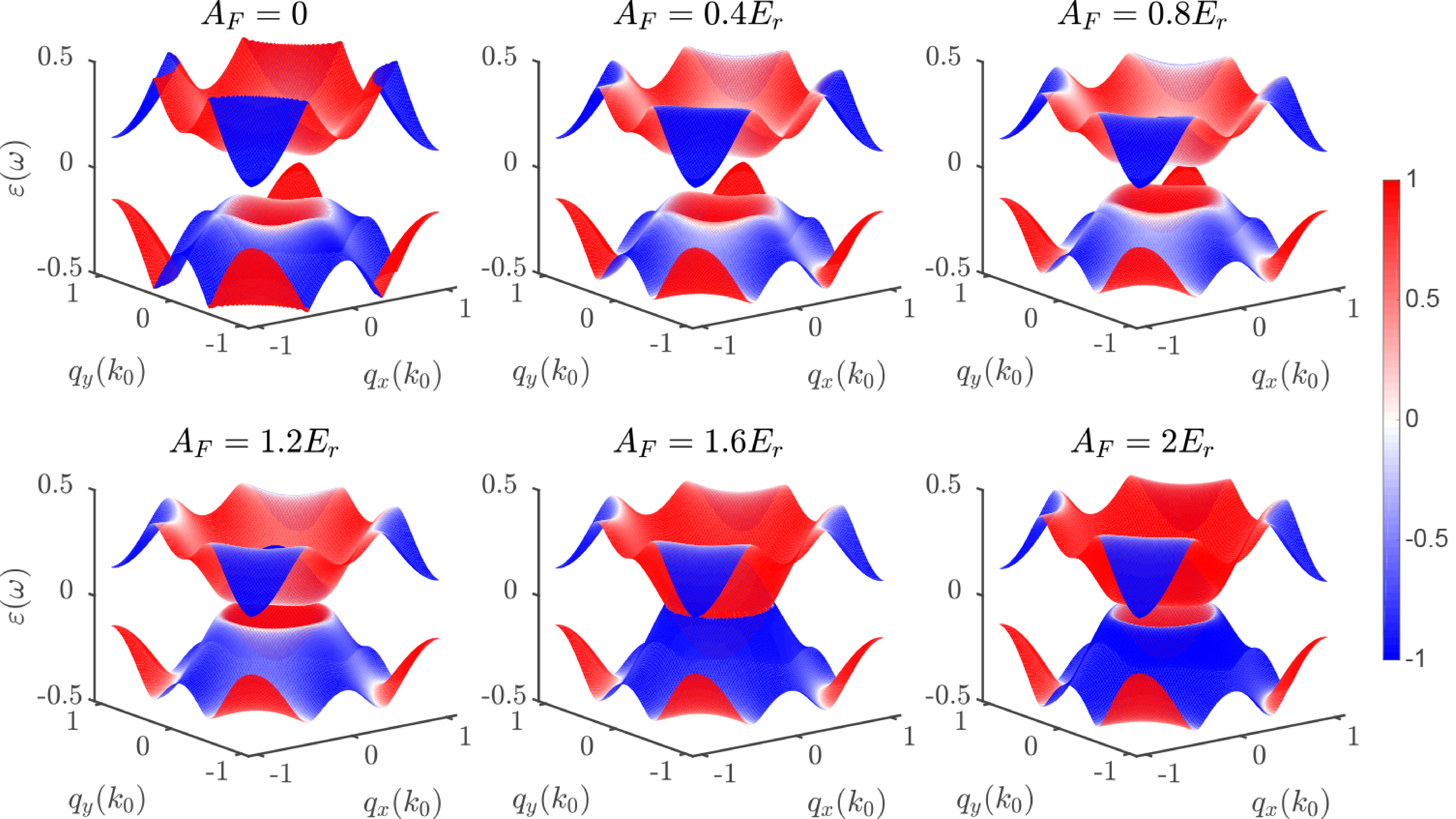}
 \caption{The Floquet bands with different driving amplitude $A_{\rm{F}}$.
  Parameters are $(V_0,~\Omega_0,~\delta_0)=(4.0,~1.0,~0.4)E_{\text{r}}$ and $T=400\mu \text s$.
  The red (blue) colour in the spin textures denotes spin up (down).
  }
  \label{FigS8}
\end{figure}

\subsection{
Numerical simulations for quenching three quantized axes
}

For periodically driven QAH model, the effective Hamiltonian $\mathcal{H}_{\rm{F}}$ can be written as
\begin{equation}\label{HF}
\mathcal{H}_{\rm{F}}=i \log(U(T))/T,
\end{equation}
where time-evolution operator $U(t)={\cal T}\exp\big[-i\int_{0}^t \mathcal{H}(\bm{q},\tau)d\tau\big]$ and ${\cal T}$ denotes the time ordering.
By diagonalizing Hamiltonian Eq.~\ref{HF} we can obtain the Floquet bands and the eigenstate $|\Psi_0(\bm{q})\rangle$. The Floquet bands with different driving amplitude $A_{\rm{F}}$ are shown in Fig.~\ref{FigS8}, manifesting that $A_{\rm{F}}$ only affects the band gap.
According to the eigenstate, the spin textures $\langle \sigma_{x,y,z} (\bm{q}) \rangle$ can be obtained by $\langle \sigma_{x,y,z} \rangle=\langle \Psi_0(\bm{q}) | \sigma_{x,y,z} |\Psi_0(\bm{q})\rangle$.
$\langle \sigma_z \rangle$ with different modulation periods $T$ are shown in Fig.~\ref{FigSpinTextureStable}.

\begin{figure}
  \includegraphics[width=0.9\linewidth]{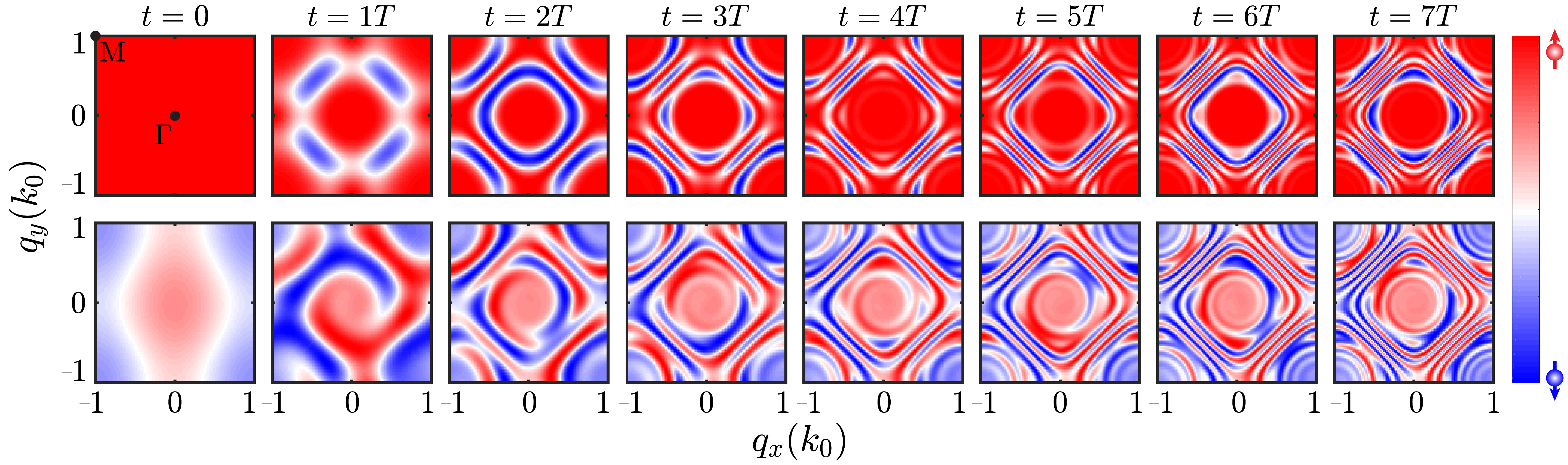}
 \caption{Evolution of the spin textures $\left\langle \sigma_z(\bm{q},t) \right\rangle$ for quenching $h_{{\rm F},z}$ (upper) and $h_{{\rm F},x}$ (lower).
  Parameters are $(V_0,~\Omega_0,~\delta_0,A_{\text{F}})=(4.0,~1.0,~0.4,~0.8)E_{\text{r}}, T=400\mu \text s$.
  $\Gamma$ and M are high symmetric momenta.
  The red (blue) colour in the spin textures denotes spin up (down).
  }
  \label{FigS9}
\end{figure}

We calculate numerically the time-evolved spin texture $\langle \sigma_z (\bm{q},t) \rangle_{i}$ along $\sigma_z$ axis by $\langle \sigma_z (\bm{q},t) \rangle_{i}=\langle \Psi(\bm{q},t)| \sigma_z | \Psi(\bm{q},t)\rangle=[N_{\uparrow}(\bm{q},t)-N_{\downarrow}(\bm{q},t)]/[N_{\uparrow}(\bm{q},t)+N_{\downarrow}(\bm{q},t)]$ with $\left| \Psi(\bm{q},t) \right\rangle=U(T)\left| \Psi_\text i \right\rangle$.
The initial state $\left| \Psi_\text i \right\rangle$ is set by the experimental conditions: For quenching $h_{\text{F},x}$, $m_x\approx -14t_0$; For quenching $h_{\text{F},y}$, $m_y\approx -14t_0$; For quenching $h_{\text{F},z}$, $\delta_0\approx -50t_0$.
The time-evolved spin textures from numerical calculations are shown in Fig.~\ref{FigS9}.
Regarding the quenching $h_{{\rm F},z}$, the spins are polarized to 1 initially since the initial detuning is far away from the resonance.
As the time $t$ increases, there emerge two rings in the spin textures due to spin flip from +1 to -1, which is the signature of BISs.
Hence, we can directly observe the rings to identify the BISs.
Regarding to quenching $h_{{\rm F},x}$, the initial spin polarization is not equal to zero because $m_{x}$ is not infinite relative to $t_0$.
With the time $t$ increases, spiral patterns form in the spin textures, which is similar to quenching $h_{\text{F},y}$.

Then, we show numerically the time-averaged spin texture $\overline{\langle \sigma_z (\bm{q}) \rangle}_{x,y,z}$ in Fig.~3 (a) of the main text by calculating the time average of $\langle \sigma_z (\bm{q},t) \rangle_{x,y,z}$.
According to Refs.~\cite{Zhang2020,Yi2019}, the time-averaged spin texture $\overline{\langle\sigma_z(\bm{q}) \rangle}_i$ involves the information of not only $h_{{\rm F},z}$ but also $h_{{\rm F},x}$ or $h_{{\rm F},y}$.
In fact, analytical calculations prove that for a fully polarized initial state (i.e., the spins are initially prepared to be polarized in the $\sigma_i$ axis for quenching $h_{{\rm F},i}$), we have $\overline{\langle\sigma_z(\bm{q}) \rangle}_i\propto-h_{{\rm F},z} (\bm{q})h_{{\rm F},i}(\bm{q})$.
That is why we use such quench processes to measure the effective spin-orbit coupling field.
For example in left panel of Fig.~3(a) in the main text, we should have $\overline{\langle\sigma_z(\bm{q})\rangle}_x\propto-h_{{\rm F},z}h_{{\rm F},x}$ under ideal conditions.
However, the results numberically finally looks a bit different because the prepared initial state cannot be fully polarized, which does not affect obtaining the dynamical field.

\end{document}